\documentclass{article}
\usepackage{amsmath,amsfonts}
\usepackage[dvipsnames]{xcolor} 
\usepackage{graphicx}
\usepackage{tikz}
\usepackage{pgfplots}
\usepackage{geometry}
\pgfplotsset{compat=1.18}

\usetikzlibrary{shapes.geometric, arrows.meta, positioning, fit, backgrounds}

\usepackage{natbib}

\usepackage{algorithm}
\usepackage{algpseudocode}

\DeclareMathOperator{\Expected}{ {\mathsf{E} }}

\DeclareMathOperator{\Exp}{\mathrm{Exp}}
\DeclareMathOperator{\Log}{\mathrm{Log}}

\newcommand{\Real}{\mathbb{R}}

\DeclareMathOperator*\argmin{\mathrm{argmin}}

\newcommand\origin{\mathfrak{o}}
\newcommand\Xspace{\mathfrak{X}}

\newcommand\Yspace{\mathfrak{Y}}

\newcommand\Hspace{\mathcal{H}}

\newcommand{\sphere}{ \mathbb{S}}

\newcommand{\sphereDist}{ \mathrm{d}_\sphere}
\newcommand{\shapeDist}{ \mathrm{d}_{\mathfrak{S}} } 

\newcommand{\YinnerProd}[1]{\langle #1 \rangle_\Yspace}
\newcommand{\HinnerProd}[1]{\langle #1 \rangle_\Hspace}
\newcommand{\norm}[1]{\lVert #1 \rVert}

\newcommand\proj{{\mathcal{P}}}

\newcommand{\adjoint}{\dagger}

\newcommand{\eval}{\mathcal{E}}

\newcommand\bounded{\mathcal{L}}
\newcommand\id{\mathrm{Id}}

\newcommand\ourMethod{\texttt{EKR}} 
\newcommand\tangentSpace{\texttt{TS}}
\newcommand\elasticTangentSpace{\texttt{ETS}}
\newcommand\kernelRidge{\texttt{KRR}}

\DeclareMathOperator*{\arginf}{arg\,inf}

\newcommand{\sphereL}{ S}

\title{Elastic kernel Ridge regression, with applications in phonetics}

\author{Beatrice Matteo\\ GSI and RISIS, GSEM, Université de Genève, Switzerland
  \and
  Almond St\"ocker\\ Institute of Mathematics, EPFL, Switzerland
  \and
  Davide Pigoli \\ 
  Department of Mathematics, King's College London, UK
  \and
  Shahin Tavakoli\footnote{email for correspondance: shahin.tavakoli@unige.ch}  \\ RISIS, GSEM, Université de Genève, Switzerland
}

\date{\today}

\begin{document}

\maketitle

\label{firstpage}

\begin{abstract}
Predicting the shapes of entire curves requires accounting for nonlinear geometry and unknown alignments between curves.
We develop elastic kernel ridge regression, a nonparametric method for planar curve responses with scalar, multivariate, or functional covariates.
Using square-root velocity representations, we formulate penalized conditional Fr\'echet mean estimation in a shape space invariant to translation, rotation, scaling, and reparametrization.
A vector-valued reproducing kernel Hilbert space provides a flexible nonlinear link through the spherical exponential map.
We use an alternating algorithm to align each observed curve to its current fitted value, and updates the regression function.
To facilitate this, we provide a new Euclidean quasi-Newton solver that exploits scale invariance of the reparametrization objective; this accelerates alignment while retaining accuracy in a numerical comparison.
Simulations demonstrate the benefits of estimating alignment within the regression and respecting spherical geometry.
Applied to vocal tract contours from a real-time magnetic resonance imaging recording, the method recovers missing frames and reconstructs shape trajectories from downsampled data.
A speech inversion proof of concept on the same recording predicts tongue shapes from acoustic features, illustrating the method's potential.
The method is implemented in the \texttt{R} package \texttt{sphereg2}.
\end{abstract}

\section{Introduction}
\label{sec:introduction}

\subsection{Motivating problem}

Elastic shape analysis \citep{srivastava2010shape} provides a geometric and statistical framework for comparing curves whose essential information lies in their shape rather than their position, orientation, scale, or parametrization. Curves are represented as points on a nonlinear, infinite-dimensional space, with distances designed to capture meaningful bending and stretching while separating shape variation from registration effects \citep{jermyn2017elastic}. This makes inference more challenging than in Euclidean spaces: correspondences between curves and population-level summaries must often be estimated jointly, while optimization over reparametrizations can be computationally demanding. These issues are central when the response is a continuous curve, which is the problem this paper addresses. 

Most regression models with shape responses focus instead on landmark-based shapes \citep{dryden2016statistical}. In that setup, shapes are defined by a fixed finite number of labeled landmarks, and analysis typically accounts for translation, scale, and rotation but resolving correspondences prior to conducting regression. Intrinsic regression models on Kendall's shape space extend linear regression to manifold-valued responses \citep{thomasfletcherGeodesicRegressionTheory2013,corneaRegressionModelsRiemannian2017a}, while Geo-FARM treats rotational alignment as a latent nuisance parameter in a landmark model \citep{huangGeoFARMGeodesicFactor2021b}. Shape-space smoothing splines provide another landmark-based approach \citep{kume2007shape}. These methods address important aspects of landmark-based shape regression, but do not consider elastic shape regression. 

Elastic regression for response curves modulo reparametrization remains a relatively small literature. Existing proposals differ in their quotient construction, alignment strategy, link flexibility, and admissible covariates \citep{steyerRegressionQuotientMetric2023,stocker2022flexible,dogra2023confounder}.
\citet{pfeuffer2026deep}, for instance, consider elastic full Procrustes means -- a different notion of shape means -- and fit them by modeling the Hermitian covariance with deep neural networks. 
A natural simplification, adopted by \citet{guo2020statistical}, is to register
the curves once to their Fr\'echet mean, map them to the tangent space at that
mean, and regress on a finite set of principal-component scores. In that approach, alignment is a preprocessing step independent of the covariates and fitted values. In contrast, our method estimates alignment within the regression.

In this paper, we develop regression methodology for responses that are full elastic planar shapes---curves in $\Real^2$ modulo translation, rotation, scaling, and reparametrization---with scalar, multivariate, or functional covariates. Our contribution is the combination three components. The response is the full elastic shape, an element of an
infinite-dimensional quotient space rather than a finite set of landmarks
with fixed correspondences. The alignment over reparametrization and rotation
is estimated jointly with the regression function, by an iterative algorithm
that registers each observation to its current fitted value rather than once
in preprocessing. The link is the nonparametric vector-valued RKHS of the
spherical kernel ridge regression of
\citet{matteo2026infinitedimensionalsphericalkernelridge}, which
accommodates scalar, multivariate and functional covariates in a single unifying formulation. 
To accelerate the alignment, which is the innermost step of every fit update, we also propose a
novel solver for the reparametrization problem, which is faster than the competing methods at no loss of accuracy
(See Appendix)

Our main application is the reconstruction of vocal tract shapes from real-time magnetic resonance imaging (rt-MRI). Earlier studies often represented articulators by a small number of landmarks, which can omit important gestures and reduce the problem to multivariate regression. Advances in real-time magnetic resonance imaging (rt-MRI) now permit measurement of vocal-tract tissues and extraction of complete articulator contours at high temporal frenquency \citep{toutiosArticulatorySynthesisBased2016,somandepalliSemanticEdgeDetection2017,limMultispeakerDatasetRaw2021}, see Figure~\ref{fig:frames_example}. These data motivate treating the contour itself as a shape-valued response. Standard functional regression for curve-valued responses \citep[see, e.g.,][]{ramsayFunctionalDataAnalysis2005} is not appropriate because the analysis should be invariant to translation, rotation, scaling, and reparametrization.

The primary we task address is temporal smoothing (Section~\ref{sec:smoothing}), where we treat acquisition time of articulator contours as the covariate and recover the shape trajectory of a recording by exploiting the temporal dependence across neighboring frames. The proposed framework can smooth articulator trajectories within a recording session, reducing measurement uncertainty. Moreover, recovering a temporally coherent sequence of shapes by imputing missing or discarded frames becomes fundamental for downstream tasks. 

We also illustrate our method on the problem of \emph{speech inversion} (Section~\ref{sec:speech inversion}). 
Speech inversion is the problem of prediction of articulator
shapes---here the tongue---from the corresponding acoustic speech signal. It is a considerably
hard problem in phonetics because the map from acoustics to articulation is nonlinear, and
ill-posed given that distinct vocal tract configurations can produce nearly the
same sound \citep{maeda1990compensatory,mrayati_distinctive_1988}. 
State-of-the-art approaches for speech inversion with complete articulator contours rely on deep networks
trained on up to several hours of speech from a single speaker \citep{azzouz2025complete,azzouz2025reconstruction,azzouz2026embeddings}. 
Our single-recording example is thus meant as a proof of concept with high-dimensional acoustic covariates rather than as a solution to the inversion problem. 
Furthermore, it shows the practical value of temporal smoothing, which completes corrupted training contours, so that no acoustic observation has to be discarded. 

The remainder of the paper is organized as follows. Section~\ref{sec:preliminaries} introduces elastic shape analysis, Sections~\ref{sec:model} and \ref{sec:computations} develop the model and its computation, and Section~\ref{sec:sims} presents simulations. Sections~\ref{sec:smoothing} and \ref{sec:speech inversion} illustrate temporal smoothing and speech inversion. 
Our methods are implemented in the \texttt{R} package \texttt{sphereg2}.

\section{Preliminaries and notation}
\label{sec:preliminaries}

\subsection{Elastic shape analysis}
\label{sec:elastic_shape_analysis}

In this paper, we consider planar curves $\beta:[0,1]\to\mathbb{R}^2$. Their square-root
velocity function (SRVF) is
\[
  q(t) := \frac{\dot{\beta}(t)}{\sqrt{\lvert\dot{\beta}(t)\rvert}},
\]
with the usual value $q(t)=0$ when $\dot\beta(t)=0$. The SRVF representation
removes translation, and its $L^2$ norm equals the square root of the curve
length. We therefore work in the Hilbert space
  $\mathcal{Q}=L^2([0,1],\mathbb{R}^2),$ with inner-product
  $\langle q_1,q_2\rangle =\int_0^1 \langle q_1(t),q_2(t)\rangle_{\mathbb{R}^2}\,dt$ and norm $\| q_1 \| = \sqrt{ \langle q_1, q_1 \rangle}$ for $q_1,q_2 \in Q$.
After restricting to unit-length curves, the pre-shape space is the unit
sphere
  $\sphere=\{q\in\mathcal{Q}:\|q\|=1\},$ 
  with distance
  $\sphereDist(p,q)=\arccos\bigl(\langle p,q\rangle\bigr).$

The SRVF representation still depends on rotation and reparametrization of
the curve, which we remove next. These are represented by the actions of
$SO(2)$ and the reparametrization group
\[
  \Gamma=\{\gamma:[0,1]\to[0,1]:\gamma(0)=0,\ \gamma(1)=1,\ \gamma
  	\text{ is a smooth diffeomorphism with }\dot\gamma>0\}.
\]
For $O\in SO(2)$ and $\gamma\in\Gamma$, define their joint action on $q\in\sphere$ by
$
  (O,\gamma)\cdot q
  :=\sqrt{\dot\gamma}\,O(q\circ\gamma),
$
see \citet[Lemma 5.1]{srivastava2016functional}.
The orbit of $q$ and the corresponding shape space are therefore
\[
  [q]=\{(O,\gamma)\cdot q:O\in SO(2),\ \gamma\in\Gamma\}
  \quad \text{and} \quad
  \mathfrak{S}=\sphere/(SO(2) \times \Gamma).
\]
The induced elastic shape distance is
\begin{equation}
  \label{eq:shapeDist}
	  \shapeDist([q_1],[q_2])
  =\inf_{O\in SO(2),\,\gamma\in\Gamma}
    \sphereDist\bigl(q_1,(O,\gamma)\cdot q_2\bigr), \quad q_1,q_2 \in \sphere.
\end{equation}
This construction endows the shape space with a proper metric, ensuring that notions such as geodesics and mean shapes can be defined, and making it possible to perform statistical analysis directly on shapes.
Computing $\shapeDist$ requires solving an optimization over $SO(2) \times \Gamma$. Practical algorithms exist for numerical alignment and statistical inference, and this paper provides a new competitive alignment method (see Section~\ref{sec:shape_alignment}). For sphere representatives $q_1,q_2\in\sphere$, when the infimum in \eqref{eq:shapeDist} is attained at
\[
  (O^*,\gamma^*)\in\argmin_{O\in SO(2),\,\gamma\in\Gamma}
  \int_0^1\left\|q_1(t)-\sqrt{\dot\gamma(t)}\,Oq_2(\gamma(t))\right\|_{\mathbb R^2}^2\,dt,
\]
we denote by
\[
  \Psi_{q_1}(q_2)(\cdot)
  :=\sqrt{\dot\gamma^*(\cdot)}\,O^*q_2(\gamma^*(\cdot))
\]
the representative of $[q_2]$ optimally aligned to $q_1$.
When an optimizer $(O^*, \gamma^*)$ is not available, $\Psi$ denotes the numerically selected
approximate alignment.

\subsection{Riemannian calculus on the sphere}

Estimation of $\mu(x)$ and the algorithms of Sections~\ref{sec:model}
and~\ref{sec:computations} require some further Riemannian structure
on $\sphere$, which we introduce here for later reference. For $p\in\sphere$, its
tangent space is
\[
  T_p\sphere=\{v\in\mathcal{Q}:\langle v,p\rangle=0\},
\]
and the Riemannian exponential map is
\[
  \Exp_p(v)=\cos(\|v\|)p+
    \sin(\|v\|)\frac{v}{\|v\|},\quad v\in T_p\sphere,
\]
where the value at $v=0$ is understood by continuity. For $q\ne-p$, $q,p \in \sphere$, the
logarithm map is
\[
  \Log_p(q)=\sphereDist(p,q)\frac{\proj_p(q)}{\|\proj_p(q)\|},
  \qquad
  \proj_p(q)=q-\langle q,p\rangle p.
\]

\subsection{Vector-valued reproducing kernels}

A vector-valued reproducing kernel Hilbert space (VVRKHS) is a Hilbert space
$\Hspace$ of functions from a covariate space $\Xspace$ to a Hilbert space
$\Yspace$ for which every pointwise evaluation map $\eval_x f=f(x), x \in \Xspace, f \in \Hspace$, is continuous. Its
operator-valued kernel satisfies
\[
  K(x,x')=\eval_x\eval_{x'}^\adjoint\in\bounded(\Yspace),
\]
where $\eval_{x'}^\adjoint$ is the adjoint of $\eval_{x'}$. 
The reproducing property is
\[
  \YinnerProd{y,f(x)}=\HinnerProd{K_x y,f},
  \qquad K_x=\eval_x^\adjoint,
  \quad x \in \Xspace, y \in \Yspace, f \in \Hspace.
\]
In this paper we use simple multiplicative operator kernels (SMO kernels)
\[
  K(x,x')=k(x,x')\id,
\]
where $k$ is a scalar RKHS kernel and $\id$ is the identity on
$\Yspace$. 

\section{Elastic shape regression}
\label{sec:model}
\subsection{Formulation of the problem}

Consider a random pair $(X,Y)$ where $X$ is the covariate (taking values in a Polish space $\Xspace$) and $Y$ is the scaled SRVF of a shape, i.e. $Y \in \sphere$. Our goal is to estimate, for every $x \in \Xspace$, a representative $\mu(x) \in \sphere$ of the conditional Fréchet mean \citep{petersenFrechetRegressionRandom2019}  of $[Y]$ given $X=x$. We tackle this through a penalized empirical risk, 
\begin{align} 
  \hat \mu &:= \argmin_{\mu} \frac{1}{n} \sum_{i=1}^n \shapeDist^2\!\big( [\mu(X_i)],\, [Y_i] \big) \;+\; J(\mu) ,
  \\ & = \argmin_{\mu} \frac{1}{n} \sum_{i=1}^n \inf_{(O, \gamma) \in SO(2) \times \Gamma} \sphereDist^2\!\big( \mu(X_i),\, \sqrt{\dot\gamma}\,O\,(Y_i \circ \gamma)\big) \;+\; J(\mu) ,
\label{eq:mu_empirical_risk}
\end{align}
where the minimum is taken over the space of functions $\mu: \Xspace \to \sphere$, and $J(\mu)$ is a measure of complexity of the function $\mu$. We shall assume throughout that the minimums are achieved (see also Section~\ref{sec:discussion}).
Solving \eqref{eq:mu_empirical_risk} is a challenging task.
One approach is to relax the problem, performing a single alignment as a preprocessing step, before performing the regression on the pre-aligned curves with fixed rotation and parametrization \citep{guo2020statistical}. Only approximately minimizing the squared elastic shape distances, 
this can work well if the optimal alignment of sampled curves $Y_i$ to $\mu(x)$ does not vary much across $x$, i.e., if the curves share their dominant features. 
However, if that is not the case, the suboptimal pre-alignment will likely translate to suboptimal conditional mean estimation, as illustrated by \citet{steyerRegressionQuotientMetric2023} for alignment over $\Gamma$ only.
We propose a more flexible strategy through an iterative optimization algorithm, alternating optimization over the alignment term $(\gamma, O)$ and over the function $\mu$, along the lines of \citet{stocker2022flexible} and \citet{steyerRegressionQuotientMetric2023}. This requires us to address the following two problems: the shape alignment problem (Section~\ref{sec:shape_alignment})
\begin{equation}\label{eq:shape_dist_opt_prob}
  \arginf_{(O, \gamma) \in SO(2) \times \Gamma} \sphereDist\!\big(q_1,\, O\,(q_2 \circ \gamma)\sqrt{\dot\gamma}\big), \quad q_1,q_2 \in \sphere,
\end{equation}
and the spherical regression problem (Section~\ref{sec:spherical_regression})
\begin{equation} \label{eq:spherical_mu_empirical_risk}
  \argmin_{\mu: \Xspace \to \sphere} \frac{1}{n} \sum_{i=1}^n \sphereDist^2\!\big( \mu(X_i) ,\, Z_i \big) \;+\;  J(\mu)
\end{equation}
where $Z_i \in \sphere$.

\subsection{Shape alignment}
\label{sec:shape_alignment}

Following \citet{huang2016riemannian}, we reformulate the optimization problem \eqref{eq:shape_dist_opt_prob} over $SO(2) \times \Gamma$ as follows. Given $q_1, q_2 \in \sphere$, we want to compute the optimal rotation $O^* \in SO(2)$ and reparametrization $\gamma^* \in \Gamma$ that minimize
\begin{align*}
H(O, \gamma) :=\int_0^1 \|q_1(t)- \sqrt{\dot \gamma(t)}Oq_2(\gamma(t))\|_{\mathbb R^2} ^2dt. 
\end{align*}

One convenient strategy is to use a block‑coordinate descent (relaxation) scheme on the product space $SO(2) \times \Gamma $, alternately minimizing with respect to $O$ and to $\gamma$, while keeping the other parameter fixed. This approach is particularly fitting for this problem because, for fixed $\gamma$, the optimal rotation has a closed-form solution given by
$\hat O=UV^T \in SO(2)$, where $U\Sigma V^T$ is the SVD of $A = \int_0^1 q_1(t)\left(q_2( \gamma(t)) \sqrt{\dot{ \gamma}(t)}\right)^T dt$.
 In principle, one would iterate the two coordinate‑descent steps until convergence of both parameters. We follow however the approach of \citet{huang2016riemannian} who advocate performing only a single outer iteration (rotation–reparametrization–rotation) and report that this yields satisfactory solutions in practice while avoiding repeated passes through the main computational bottleneck of optimization over $\Gamma$.
 For fixed $O$ the problem reduces to an optimization over the group of reparametrizations, which has no closed-form solution. The standard approach is dynamic programming \citep[DP;][Appendix~B]{srivastava2016functional}, which restricts $\gamma$ to monotone piecewise-linear paths on a grid over $[0,1] \times [0,1]$. It is widely used and finds the global minimum over such paths, but the accuracy is capped by the spacing of the grid, and the cost grows quadratically with the number of grid points. Since the alignment will be the innermost step of every model fit, this is prohibitive for our purposes. \citet{huang2016riemannian} instead cast the problem as a Riemannian optimization over the unit sphere $\sphereL$ of $L^2([0,1],\mathbb{R})$, on which a reparametrization is represented by $l = \sqrt{\dot\gamma}$, and propose a steepest descent and a Broyden--Fletcher--Goldfarb--Shanno (BFGS) scheme on $\sphereL$.

We build on the latter formulation and its Riemannian gradient to propose a new Euclidean optimizer that achieves similar alignment accuracy with lower computational time. 
Writing $\gamma_l(t) = \int_0^t l^2(s)\,ds \big/ \int_0^1 l^2(s)\,ds$, for $l \in L^2([0,1],\mathbb{R})$, the cost $H(\gamma_l, O)$ is invariant under the rescaling $l \mapsto c\,l$ for every $c>0$, that is, positively homogeneous of degree zero. 
The constraint $\int_0^1 l^2(t) dt=1$ may therefore be dropped and the reparametrization objective can be minimized over the whole of $L^2([0,1],\mathbb{R})$ by an Euclidean method. 
We choose to use the limited-memory BFGS (LBFGS) algorithm of \citet{byrd1995limited}. 
Being gradient-based, it is guaranteed to converge only to a local minimum of the cost, but in the comparison of the Appendix is nevertheless more accurate than DP on every replication. 
Moreover, one can show that the gradient of $H(\gamma_l, O)$ coincides with the Riemannian gradient, up to renormalization, so that a single gradient implementation serves both Euclidean and Riemannian gradient-based formulations. 
A similar approach can be found in the deep-learning literature \citep{arora2019auto}

For more details on the two schemes of \citet{huang2016riemannian}, of which we provide the first \texttt{R} implementation in \texttt{sphereg2}, the derivation of the scheme just described, a review of the other approaches that have been proposed for the reparametrization problem, and a numerical comparison on a real vocal tract contour showing our scheme is the fastest while retaining accuray, see the Appendix.

\subsection{Spherical regression}
\label{sec:spherical_regression}

The optimization problem \eqref{eq:spherical_mu_empirical_risk} is a challenging non-linear and non-convex optimization problem. We make it tractable by
using the method of \citet{matteo2026infinitedimensionalsphericalkernelridge}.
We write $\mu(x) = \Exp_\origin f(x)$ for some point $\origin \in \sphere$ and some function $f$ belonging to a VVRKHS $\Hspace$ of functions $f: \Xspace \to T_\origin \sphere$ with operator kernel $K(x,x') = k(x,x')\id$. Assuming $\origin$ is fixed, the resulting penalized empirical risk is given by
\begin{equation} \label{eq:penalized_empirical_risk_vvrkhs}
   \frac{1}{n} \sum_{i=1}^n \sphereDist^2\!\big( \Exp_{\origin} f(X_i),\, Z_i \big) \;+\; \lambda_n^2 \|f\|_{\Hspace}^2.
\end{equation}
For $\lambda_n > 0$, a minimizer of \eqref{eq:penalized_empirical_risk_vvrkhs} always exists and is of the form
$\hat f = \sum_{i=1}^n k(\cdot, x_i) \xi_i$ where $\xi_i \in T_\origin \sphere$ \citep[][Theorem~4.1]{matteo2026infinitedimensionalsphericalkernelridge}. The estimator $\hat \mu = \Exp_\origin \circ \hat f$ is then shown to be a consistent estimator of the conditional Fréchet mean
\[
    \mu(x) = \arg\min_{q \in \sphere} \Expected\bigl( \sphereDist^2(q, Z) \mid X=x \bigr),
\]
which is well defined under a bounded conditional support condition on the law of $Z\mid X$ \citep[for details see][]{matteo2026infinitedimensionalsphericalkernelridge}. 

\section{Computational aspects}
\label{sec:computations}

The optimization strategy used here builds directly on the algorithm of \citet{matteo2026infinitedimensionalsphericalkernelridge}, as implemented in their companion \texttt{R} package. In that setting, all computations were carried out in a low-rank representation, obtained by projecting the kernel and response Gram matrices onto their leading eigenspaces, and BFGS was applied to the resulting low-dimensional parametrization. In the present elastic shape framework, however, aligning the predicted curve at each BFGS iteration with the observed shape requires an optimization over $(O, \gamma )\in SO(2) \times \Gamma$, which must be performed in the original SRVF space.

To reconcile these two requirements, we proceed as follows. Our algorithm proceeds through the following steps at each iteration:
\begin{itemize}
    \item We map the current low-rank representation of the fitted responses back to the full SRVF space, reconstructing the corresponding curves in $L^2([0,1],\mathbb{R}^2)$, represented in this implementation by their evaluation on a fixed grid. We then perform the elastic alignment step, computing the optimal reparametrization and rotation that align the predicted SRVF with the observed one, as described in Section~\ref{sec:shape_alignment}, see Algorithm~\ref{alg:align_step}.
    \item The aligned responses are subsequently re-projected onto low-rank eigenspaces, see Algorithm~\ref{alg:reduced_basis}.
    \item These low rank representations of the aligned responses are used to update the regression function, see Algorithm~\ref{alg:bfgs_update}. 
\end{itemize}
The overall procedure is given in Algorithm~\ref{alg:elastic_sphere_bfgs}.

\begin{algorithm}[p]
  \caption{\textsc{AlignResponses}: alignment step}
  \label{alg:align_step}
  \begin{algorithmic}[1]
    \Function{AlignResponses}{$f,\, (X_i,Y_i)_{i=1}^n,\, \origin$}
      \Require Regression function $f$, training pairs $(X_i,Y_i)_{i=1}^n$, base point $\origin \in \sphere$
      \Ensure Alignments $(O_i, \gamma_i)_{i=1}^n$ and aligned responses $(\tilde Y_i)_{i=1}^n$
      \For{$i = 1,\dots,n$}
        \State $\hat Y_i(t) \gets \Exp_{\origin}\bigl(f(X_i)\bigr)(t)$
          \Comment{predicted SRVF on the time grid}
        \State $(O_i, \gamma_i) \gets \displaystyle\argmin_{O \in SO(2), \, \gamma \in \Gamma}
            \int_0^1 \bigl\|\hat Y_i(t) - \sqrt{\dot\gamma(t)}\,O\,(Y_i \circ \gamma)(t)\bigr\|_{\mathbb{R}^2}^2 \, dt$
          \Comment{coordinate descent over $(O, \gamma)$}
        \State $\tilde Y_i \gets \sqrt{\dot\gamma_i}\;O_i\,(Y_i \circ \gamma_i)$
          \Comment{aligned response}
      \EndFor
      \State \Return $(O_i, \gamma_i,\tilde Y_i)_{i=1}^n$
    \EndFunction
  \end{algorithmic}
\end{algorithm}

\begin{algorithm}[p]
  \caption{\textsc{ReducedBasis}: low-rank response projection}
  \label{alg:reduced_basis}
  \begin{algorithmic}[1]
    \Function{ReducedBasis}{$(\tilde Y_i)_{i=1}^n,\, m_y$}
      \Require Aligned responses $(\tilde Y_i)_{i=1}^n$, target rank $m_y$
      \Ensure Low-dimensional representations $(\mathbf{z}_i)_{i=1}^n \subset \mathbb{R}^{m_y}$
      \State $\mathbf{G} \gets \bigl[\langle \tilde Y_i, \tilde Y_j\rangle\bigr]_{i,j=1}^n$
        \Comment{aligned response Gram matrix}
      \State Compute leading $m_y$ eigenvectors of $\mathbf{G}$, stack them as columns of an $n \times m_y$ matrix $M$, and set $\mathbf{z}_i \gets \text{row}_i(M) \in \mathbb{R}^{m_y}$, $i=1,\dots,n$.
      \State \Return $(\mathbf{z}_i)_{i=1}^n$
    \EndFunction
  \end{algorithmic}
\end{algorithm}

\begin{algorithm}[p]
  \caption{\textsc{BFGSUpdate}: regression update in reduced space}
  \label{alg:bfgs_update}
  \begin{algorithmic}[1]
    \Function{BFGSUpdate}{$\boldsymbol{\xi},\, (X_i, \tilde Y_i)_{i=1}^n,\, \lambda_n$}
      \State Let $f_{\boldsymbol{\xi}} := \sum_{i=1}^n K(\cdot, X_i)\xi_i$ 
      \Require Current coefficients $\boldsymbol{\xi}$, covariate and aligned responses pairs $(X_i, \tilde Y_i)_{i=1}^n$, regularization $\lambda_n$
      \Ensure Updated coefficients $\boldsymbol{\xi}'$ and regression function $f_{\boldsymbol{\xi}'}$
      \State Define the reduced empirical risk
        \[
          R_n(\boldsymbol{\xi})
          = \frac1n \sum_{i=1}^n
            \sphereDist^2\!\bigl(
              \Exp_{\origin} f_{\boldsymbol{\xi}}(X_i),\, \tilde Y_i
            \bigr)
            + \lambda_n \|f_{\boldsymbol{\xi}}\|_{\Hspace}^2
        \]
      \State Compute the gradient $\nabla_{\boldsymbol{\xi}} R_n(\boldsymbol{\xi})$ \Comment{see \citet{matteo2026infinitedimensionalsphericalkernelridge}}
      \State Update $\boldsymbol{\xi}$ with BFGS using $R_n$ and $\nabla_{\boldsymbol{\xi}} R_n$ to obtain $\boldsymbol{\xi}'$ and $f_{\boldsymbol{\xi}'}$
      \State \Return $\boldsymbol{\xi}', f_{\boldsymbol{\xi}'}$
    \EndFunction
  \end{algorithmic}
\end{algorithm}

\begin{algorithm}[p]
  \caption{Elastic spherical regression with embedded alignment}
  \label{alg:elastic_sphere_bfgs}
  \begin{algorithmic}[1]
    \Require Training data $(X_i,Y_i)_{i=1}^n$ with $X_i \in \Xspace$, SRVF responses
      $Y_i \in \sphere \subset L^2([0,1],\mathbb{R}^2)$ (discretized on a fixed time grid),
      kernel $k$, regularization parameter $\lambda_n$, base point $\origin \in \sphere$

    \State $[\mathbf{K}]_{i,j} \gets k(X_i, X_j)$
      \Comment{kernel Gram matrix, shared across iterations}
    \State Compute leading $m_k$ eigenvectors of $\mathbf{K}$ to obtain basis $g_1,\dots,g_{m_k}$
      \Comment{reduced predictor space, computed once}

    \State $f^{(0)} = f_{\boldsymbol{\xi}^{(0)}} \gets$ tangent-space regression estimate
    \State $\gamma_i^{(0)} \gets \mathrm{id}$, $O_i^{(0)} \gets I_2$, for $i=1,\dots,n$

    \For{$k = 0,1,2,\dots$ until convergence}
      \State $(O_i^{(k+1)},\gamma_i^{(k+1)},\tilde Y_i^{(k+1)})_{i=1}^n \gets$
        \Call{AlignResponses}{$f^{(k)},\, (X_i,Y_i)_{i=1}^n,\, \origin$}
      \State $(\mathbf{z}_i^{(k+1)})_{i=1}^n \gets$
        \Call{ReducedBasis}{$(\tilde Y_i^{(k+1)})_{i=1}^n,\, m_y$}
      \State $(\boldsymbol{\xi}^{(k+1)}, f^{(k+1)}) \gets$
        \Call{BFGSUpdate}{$\boldsymbol{\xi}^{(k)},\, (X_i, \tilde Y_i^{(k+1)})_{i=1}^n,\, \lambda_n$}
    \EndFor

    \State \textbf{Output:} final estimator $f^{(\hat k)}$ and alignments $(O_i^{(\hat k)}, \gamma_i^{(\hat k)})_{i=1}^n$
  \end{algorithmic}
\end{algorithm}

\begin{figure}[H]
  \centering
  \begin{tikzpicture}[
      box/.style={
        rectangle, rounded corners, draw=black, thick,
        minimum width=6.2cm, minimum height=1cm,
        align=center, font=\small
      },
      setupbox/.style={
        rectangle, rounded corners, draw=black!60, thick, dashed,
        minimum width=5.6cm, minimum height=0.9cm,
        align=center, font=\small
      },
      decision/.style={
        diamond, draw=black, thick, aspect=2.2,
        align=center, font=\small, inner sep=1pt
      },
      arrow/.style={-{Latex[length=2.5mm]}, thick},
      node distance=0.9cm and 1.4cm
    ]

    \node[setupbox] (precompute) {\textbf{Precompute}\\[1pt] Kernel Gram matrix and reduced predictor basis};
    \node[setupbox, below=of precompute] (init) {\textbf{Initialize}\\[1pt] Regression function and alignments};

    \node[box, below=2.1cm of init] (align) {Align fitted responses\\[1pt] (Algorithm~\ref{alg:align_step})};
    \node[box, below=of align] (reduce) {Reduce dimension\\[1pt] (Algorithm~\ref{alg:reduced_basis})};
    \node[box, below=of reduce] (regress) {Compute Spherical Regression (BFGS update)\\[1pt] (Algorithm~\ref{alg:bfgs_update})};

    \node[decision, below=1.1cm of regress] (converged) {Converged?};
    \node[box, below=1.1cm of converged] (output) {\textbf{Output}\\[1pt] final estimator and alignments };

    \begin{scope}[on background layer]
      \node[fill=black!5, rounded corners, fit=(precompute)(init), inner sep=8pt, label={[font=\small\itshape]above:Setup (once)}] (setupfit) {};
      \node[fill=blue!5, rounded corners, fit=(align)(reduce)(regress), inner sep=10pt, label={[font=\small\itshape]:\textbf{Iterate over $k$} \phantom{asdfasdfasfdjasdfasdfasdf}}] (loopfit) {};
    \end{scope}

    \draw[arrow] (precompute) -- (init);
    \draw[arrow] (init) -- (align);
    \draw[arrow] (align) -- (reduce);
    \draw[arrow] (reduce) -- (regress);
    \draw[arrow] (regress) -- (converged);
    \draw[arrow] (converged) -- (output) node[midway, right, font=\small] {yes};

    \draw[arrow] (converged.west)
      -- ++(-3.1cm,0)
      node[midway, above, font=\small] {no}
      |- (align.west);

  \end{tikzpicture}
  \caption{Conceptual overview of Algorithm~\ref{alg:elastic_sphere_bfgs}: a one-time setup phase followed by an iterative loop alternating alignment, low-rank reduction, and BFGS update of the spherical regression, until convergence.}
  \label{fig:elastic_sphere_reg_diagram}
\end{figure}
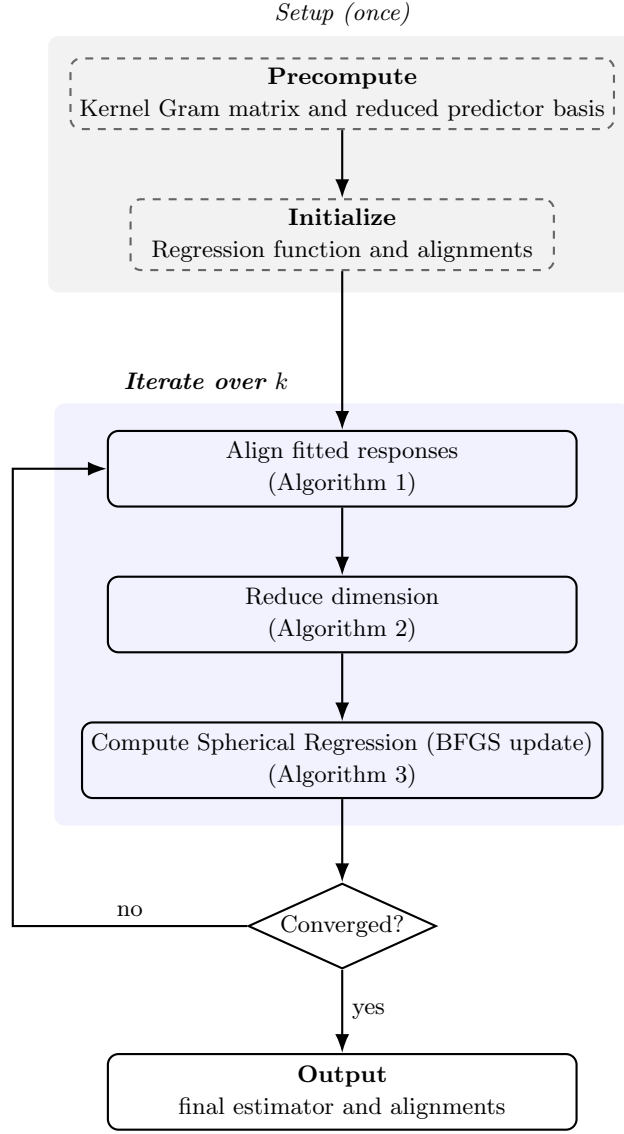

\section{Simulation study}

\label{sec:sims}
\subsection{Overview}

\begin{figure}[ht]
  \centering
  \includegraphics[width=1\textwidth]{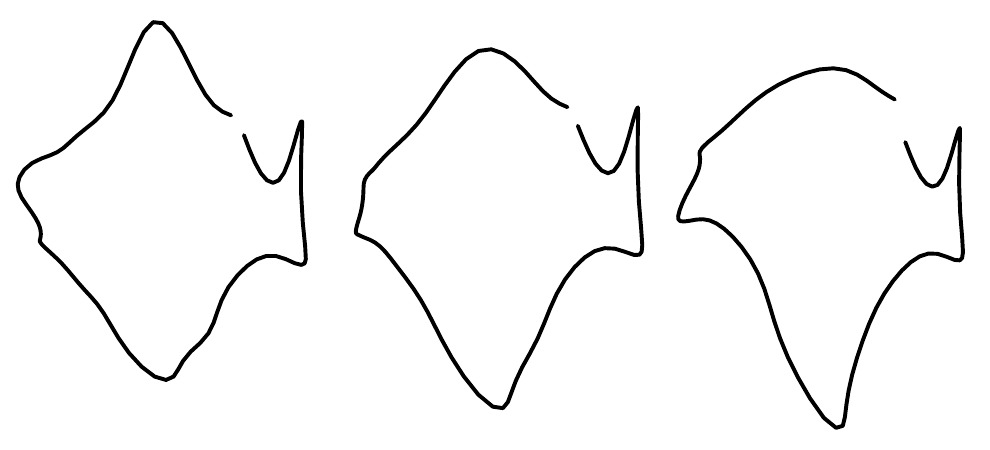}
  \caption{The true conditional mean shape $\mu(x)$ of the simulation at
  $x = -0.11$, $-0.01$ and $0.10$ (left to right): the shape evolves with the
  covariate $x$, which plays the role of time. The trajectory is built from
  six hand-drawn fish outlines represented by their SRVFs: two elastic
  spherical regressions, each fitted to three of them placed at
  $x = -1, 0, 1$, give the trajectories $\eta_0$ and $\eta_1$ of
  Section~\ref{sec:simul_setup}, and $\mu(x)$ is the midpoint of the geodesic
  joining $\eta_0(x)$ to the aligned $\eta_1(x)$. The responses are then
  drawn on that geodesic around $\mu(x)$ (See the Appendix).}
  \label{fig:fish_evolution}
\end{figure}

We use simulations to assess two aspects of the proposed estimator: its
finite-sample behavior as the sample size increases, and the contributions of
elastic alignment and intrinsic spherical geometry. We compare elastic
spherical regression with methods that either pre-align the observations once,
fit in a tangent space, or ignore the SRVF representation. The covariate
$x\in\Xspace=[-1,1]$ plays the role of time and controls the evolution of a
shape (Figure~\ref{fig:fish_evolution}).

\subsection{Simulation setup} \label{sec:simul_setup}
\begin{figure}
  \begin{center}
\begin{tikzpicture}
  \begin{axis}[
      hide axis,
      clip=false,
      domain=-1:1,
      samples=100,
      width=12cm,
      height=7.5cm,
      xmin=-1, xmax=1,
      ymin=-2.4, ymax=2.4
  ]
    \addplot[thick, black] { 1.6 + 0.3*sin(180*x) + 0.1*cos(360*x)};
    \addplot[thick, black] { 0.0 + 0.3*sin(180*x) + 0.1*cos(360*x)};
    \addplot[thick, black] {-1.6 + 0.3*sin(180*x) + 0.1*cos(360*x)};
    \node[right] at (axis cs:1,1.7)  {$\Psi_{\eta_0}(\eta_1)$};
    \node[right] at (axis cs:1,0.1)  {$\mu$};
    \node[right] at (axis cs:1,-1.5) {$\eta_0$};

    \fill (axis cs:-0.2,1.4546)  circle (2pt);
    \fill (axis cs:-0.2,-0.1454) circle (2pt);
    \fill (axis cs:-0.2,-0.7854) circle (2pt);   
    \fill (axis cs:-0.2,-1.7454) circle (2pt);
    \draw[dashed] (axis cs:-0.2,1.4546) -- (axis cs:-0.2,-2.2);
    \node[above left] at (axis cs:-0.2,1.4546)  {$\Psi_{\eta_0}(\eta_1)(x)$};
    \node[above left] at (axis cs:-0.2,-0.1454) {$\mu(x)$};
    \node[below left] at (axis cs:-0.2,-1.7454) {$\eta_0(x)$};
    \node[right]      at (axis cs:-0.2,-0.7854) {$Y$ for $\tau = 0.7$};

    \draw (axis cs:-1,-2.2) -- (axis cs:1,-2.2);
    \draw (axis cs:-1,-2.13) -- (axis cs:-1,-2.27);
    \draw (axis cs:1,-2.13) -- (axis cs:1,-2.27);
    \draw (axis cs:-0.2,-2.13) -- (axis cs:-0.2,-2.27);
    \node[below] at (axis cs:-1,-2.27) {$-1$};
    \node[below] at (axis cs:1,-2.27) {$1$};
    \node[below] at (axis cs:-0.2,-2.27) {$x$};
  \end{axis}
\end{tikzpicture}
\end{center}
  \caption{Illustration of the data-generating mechanism of the simulation.
  For each covariate value $x$, the trajectory $\eta_0(x)$ and the aligned
  trajectory $\Psi_{\eta_0}(\eta_1)(x)$ span a geodesic of $\sphere$, whose
  midpoint $\mu(x)$ is the true conditional mean. A response $Y$ at $x$ is
  drawn on this geodesic at the position $\tau \sim
  \mathrm{Unif}([0.5-\rho, 0.5+\rho])$ of~\eqref{eq:simul_data_generation},
  shown here for $\tau = 0.7$, so that $\rho$ controls how far the responses
  spread around $\mu(x)$. The vertical direction stands for the position
  along the geodesic and is not a coordinate.}
  \label{fig:idea_simulation}
\end{figure}
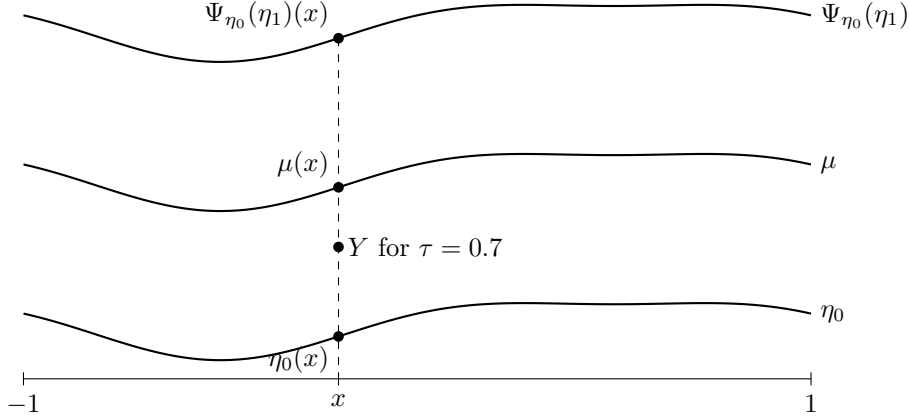

The data-generating mechanism constructs two smooth shape trajectories,
$\eta_0(x) \in \sphere$ and $\eta_1(x)\in \sphere$, $x \in [-1,1]$, and generates responses around their elastic midpoint.
This gives direct control over both the conditional mean \citep[unlike approximate simulations, e.g.,][]{steyer2023elastic} and the amount of
shape variation \citep[unlike bootstrap-type simulations, e.g.,][]{stocker2026elasticfullproc} while preserving the alignment structure of the quotient
space. For each $x \in [-1,1]$, let $\Psi_{\eta_0}(\eta_1)(x) \in \sphere$ denote the optimally aligned
version of $\eta_1(x)$ relative to $\eta_0(x)$. Alignment is preserved along
the geodesic joining the two shapes \citep[][Lemma 4]{steyerRegressionQuotientMetric2023},
so their midpoint is
\begin{align*}
     \mu(x) = \Exp_{\eta_0(x)}\!\left\{\frac{1}{2}
     \Log_{\eta_0(x)}\big(\Psi_{\eta_0}(\eta_1)(x)\big)\right\}.
\end{align*}
The response is then sampled along the geodesic joining $\eta_0(x)$ and $\eta_1(x)$,
\begin{align}\label{eq:simul_data_generation}
      Y \mid \tau &= \Exp_{\mu(x)} \big\{ \tau \Log_{\mu(x)}( \eta_0(x)) + (1-\tau) \Log_{\mu(x)}( \eta_1(x) ) \big\}, \\
      \tau &\sim \text{Unif}([0.5-\rho, 0.5 + \rho]), \nonumber
\end{align}
where $\rho\in[0,0.5]$ controls the perturbation magnitude. Since $\mu(x)$
is the midpoint, $\rho=0$ gives no perturbation, while $\rho=0.5$ samples
uniformly over the full segment between the two endpoints. The construction is
illustrated in Figure~\ref{fig:idea_simulation}.

 \subsubsection{Construction of $\eta_0$ and $\eta_1$}
We start from six SRVF shapes discretized on $100$ points (see the Appendix).
Using the first three and the last three shapes separately, we fit two elastic
spherical regression models at $x_1=-1$, $x_2=0$, and $x_3=1$. Both use a Gaussian radial basis function (GRBF)
kernel with bandwidth $\sigma=1$, the median distance between the covariates,
and the same template $\origin$, which is distinct from the six generating
shapes. The fitted trajectories are
\begin{equation*}
    \eta_0 (\cdot) = \Exp_\origin \left(\sum_{i = 1}^3 k (x_i, \cdot)\xi_i^{(1)}\right) \quad \text{and} \quad
    \eta_1 (\cdot) = \Exp_\origin \left(\sum_{i = 1}^3 k (x_i, \cdot)\xi_i^{(2)}\right).
\end{equation*}
For a sample of size $n$, the covariates $X_1,\ldots,X_n$ are equispaced on
$\Xspace$, and the responses are drawn independently from~\eqref{eq:simul_data_generation}.

 \subsubsection{Competitor models}

The proposed method, \ourMethod, updates the elastic
alignment during fitting and performs regression on the sphere. We are going to compare it with three existing methods: a tangent space regression method (\tangentSpace), an elastic tangent space regression method (\elasticTangentSpace) and a kernel Ridge regression method on the original curves (\kernelRidge). These alternatives are chosen to explore the roles of representation,
alignment, and geometry in the procedure.

The
	\tangentSpace\ method performs tangent space regression: it first aligns every observation once to the Fréchet mean and then
fits a VVRKHS regression to the projected SRVFs,
$\tilde Y_i=\Log_\origin Y_i$, using a GRBF kernel
\citep{kadri2016operator}. Thus it tests the effect of replacing joint
alignment with one-time pre-alignment.

The \elasticTangentSpace\ method performs elastic tangent space regression, and improves on \tangentSpace\ by updating the elastic alignment during
fitting, but still performs the regression in the single tangent space $T_\origin \sphere$.
At iteration $s$, it updates each unaligned response $Y_i$ into an aligned response $Y^{(s+1)}_i$ according to
 \begin{align*}
   Y^{(s+1)}_{i} = \Psi_{\hat Y^{(s)}_{i} }(Y_i ),
 \end{align*}
where $\hat Y^{(s)}_{i}$ is the model's shape prediction at $X_i$. The
alternation between fitting the regression and aligning the curves runs for at most 10 iterations. In both elastic methods (\ourMethod\ and \elasticTangentSpace), the
alignment step uses the quasi-Newton method \texttt{LBFGS} described in
Section~\ref{sec:shape_alignment} and the Appendix.

Finally, the \kernelRidge\ method transforms the pre-aligned SRVFs back to curves,
translates them to the origin, rescales them to unit length, and fits a kernel
ridge regression directly to their coordinate functions in
$L^2([0,1],\mathbb{R}^2)$. It therefore ignores both the SRVF representation
and iterative elastic alignment.

 \subsubsection{Simulation parameters and performance measure}
We consider sample sizes $n\in\{20,50,100\}$ and perturbation levels
$\rho\in\{1/6,1/3,1/2\}$, with $B=50$ repetitions for each of the nine
combinations. All methods use the same GRBF bandwidth, $\sigma=1$. Only the
regularization parameter is tuned, by five-fold cross-validation over
$\lambda\in\{10^{-5},10^{-3.5},10^{-2},10^{-0.5},10\}$. For the elastic
methods, cross-validation uses elastic distance after alignment, whereas
\tangentSpace\ and \kernelRidge\ use their corresponding loss on the
pre-aligned data.

For method $m$ and repetition $b$, let $\hat\mu_{m,b}$ denote the fitted
conditional mean. We measure performance by its average elastic distance from
the true conditional mean,
\begin{align*}
  \widehat{\text{err}}_m = \frac{1}{B} \sum_{b = 1}^B \Expected_X
  [\shapeDist( \mu (X), \hat \mu_{m,b} (X))],
\end{align*}
where the expectation is approximated on a grid of 200 equispaced points on $[-1,1]$.

\subsection{Simulation results}

\begin{figure}[ht]
  \centering
  \input{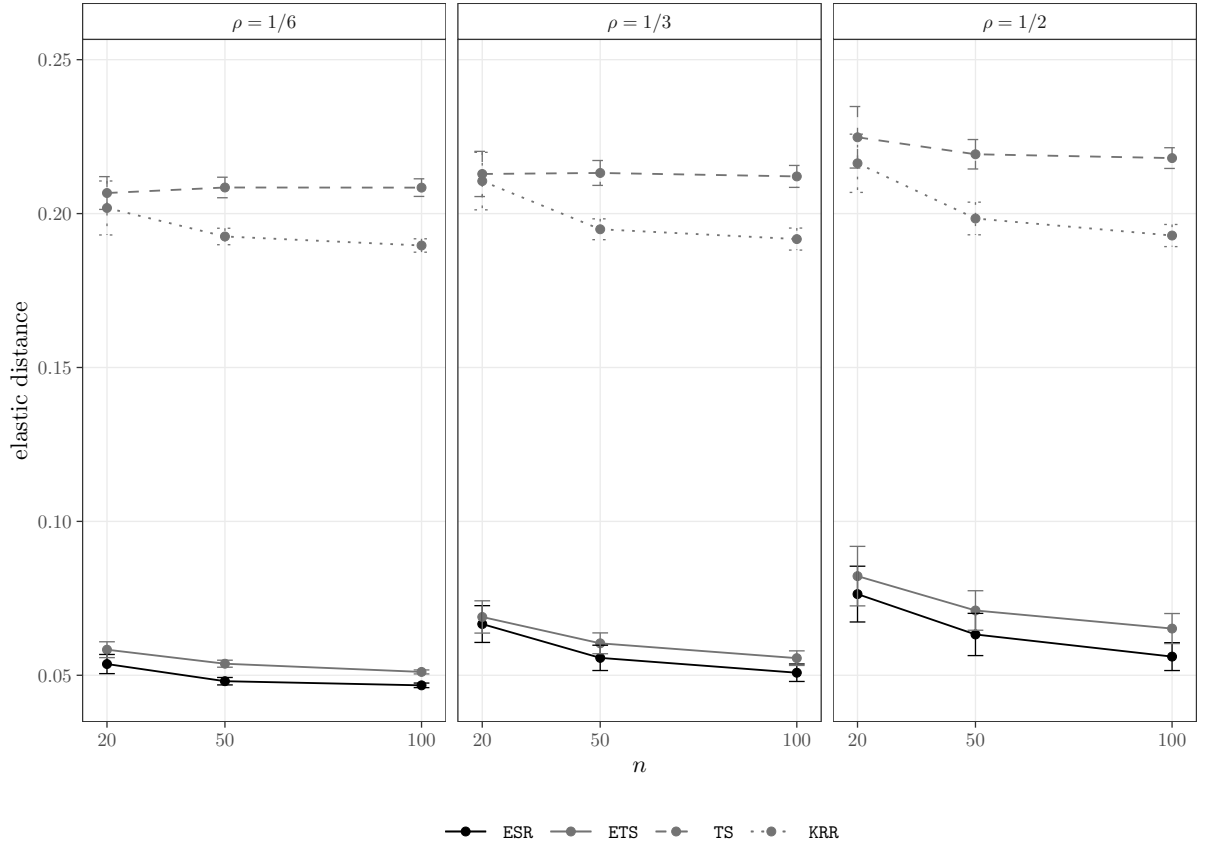}
  \caption{Mean elastic distance between the estimated and the true conditional mean shape, against the sample size $n$, for perturbation levels $\rho = 1/6, 1/3, 1/2$ (panels). Points are averages over $B = 50$ repetitions, error bars $\pm 1$ standard deviation across repetitions.}
  \label{fig:simulation_results}
\end{figure}

Figure~\ref{fig:simulation_results} reports $\widehat{\text{err}}_m$ for
each method, sample size, and perturbation level and the error bars show one standard
deviation across repetitions.

The proposed method, \ourMethod, has the smallest error in all nine settings.
For every method, error decreases as $n$ increases. Error increases with increasing
perturbation level $\rho$. Larger perturbations also produce greater variation
across repetitions.

The two methods that do not update the elastic alignment (\tangentSpace\ and \kernelRidge) perform poorly and
show little improvement with $n$. This indicates that their error is dominated by the residual misalignment
left by a single pre-alignment to the Fréchet mean rather than by estimation
variance, which more data would reduce.

The comparison with \elasticTangentSpace\ isolates intrinsic geometry from
alignment. Both methods update the elastic alignment, but
\elasticTangentSpace\ performs regression in the tangent space at $\origin$,
whereas \ourMethod\ stays on the sphere. \ourMethod\ is better in every
setting, and the gap widens with both $n$ and $\rho$, when the responses spread
farther from $\origin$ and tangent-space linearization becomes less accurate.

\section{Temporal smoothing of vocal tract shapes}

\label{sec:smoothing}
\subsection{Overview and motivation}

The data consist of a 14 second long real-time MRI video of a speech recording, from which the contours of the vocal tract articulators are extracted frame by frame by the frequency-domain segmentation of \citet{Bresch2009Region} on the corpus of \citet{somandepalliSemanticEdgeDetection2017, toutiosArticulatorySynthesisBased2016}. 
Treated as a single open planar curve, this sequence of shapes is an imperfect measurement of the underlying articulatory motion: the automated segmentation is subject to error (due to limited spatial resolution), and individual frames may be noisy, unreliable, or effectively missing.

We frame the reconstruction of the continuous-time trajectory underlying the observed sequences of articular contours as a shape regression problem with covariate space $\Xspace = [-1,1]$ (i.e., rescaled time), and the fitted regression function estimating the vocal tract shape at every time point. We call this temporal smoothing. This framing serves three purposes. First, it denoises the observed trajectory, since each fitted shape borrows strength from neighboring frames while contour extraction artifacts are smoothed out. Second, it imputes the shapes of frames that are missing or discarded as unreliable, yielding complete trajectories for downstream analyses, in particular for speech inversion (Section~\ref{sec:speech inversion}), where discarding a frame means discarding the corresponding acoustic observation. Third, because the covariate is continuous, the fitted model can be evaluated at arbitrary time points, allowing shape sampling at a slower frame rate than the acquisition rate: the recordings analyzed here have a comparatively high temporal resolution of about $83$ frames per second, but rates around $50$ frames per second are common in other rt-MRI corpora \citep{azzouz2025complete}, and downsampling (or even upsampling) would enable researchers to apply various rtMRI dataset to their existing pipeline, allowing also performance comparison across various datasets. 

\subsection{From segmentation masks to shape data}
\label{sec:contour_ordering}

For each frame, the segmentation algorithm returns a pixel-wise label mask on an $84\times 84$ grid, assigning each pixel to one of the articulator classes; the pixels carrying a given label form an unordered planar point set, and recovering their order along each articulator is a necessary pre-processing step. We do so with a graph-theoretic procedure, taking the labeled pixels as the vertices of a complete graph with Euclidean edge weights and reading the ordering off the diameter of its minimum spanning tree \citep{kruskal1956shortest}, implemented with the \texttt{igraph} package \citep{csardi2006igraph} in \texttt{sphereg2}; full details are given in the Appendix. The ordered contours are then oriented consistently, concatenated into a single open curve, smoothed with a smoothing spline, evaluated on a regular grid of $100$ points, and mapped to the SRVF representation of Section~\ref{sec:preliminaries}.

Of the $1161$ frames of the recording, $46$ carried no segmentation output at all; of the remaining $1115$, our ordering algorithm flags $105$ as unusable, in the sense that the ordered contour omits more than five of the pixels the segmentation labeled on at least one articulator, leaving $1010$ frames for the analysis.

\subsection{Model fitting and results}
We analyze a recording acquired at $83.28$ frames per second and assess performance, on the $1010$ frames of Section~\ref{sec:contour_ordering}, under two hold-out designs. The \emph{missing-frame recovery} design removes one chunk of five consecutive frames from every window of twenty, leaving gaps of $60$ ms. The \emph{low-frame-rate} design keeps only the first of every four consecutive frames for training and the third for testing, discarding the other two, so that the trajectory is reconstructed from a sequence acquired four times more slowly than the recording, with every held-out frame two frames from the nearest training one. We fit \ourMethod\ with a GRBF kernel, its bandwidth and regularization parameter selected separately for each design by four-fold cross-validation on the training frames, with elastic distance $\shapeDist$ as metric.

We compare \ourMethod\ with two regression models for elastic shape responses---the quotient linear regression of \citet{steyerRegressionQuotientMetric2023} (\texttt{STEYER}), linear in the covariate and fitted with the authors' \texttt{elasdics} package \citep{steyer2023elasdics}, and the tangent-space principal component regression of \citet{guo2020statistical} (\texttt{GUO}), which registers curves once to their Fréchet mean before an ordinary linear regression on the first $20$ tangent PC scores ($>99\%$ of the tangent variance), implemented by us as no code accompanies that paper---and with the Fréchet mean of the training shapes as covariate-free baseline (\texttt{FM}). 

Figure~\ref{fig:boxplots_smoothing} displays, under both designs, the distribution of the elastic distance $\shapeDist$ between each held-out shape and its prediction. \ourMethod\ improves on the baseline, \texttt{GUO} is level with it, and \texttt{STEYER} is above it.

\begin{figure}[t]
\centering
  \includegraphics[width=.70\textwidth]{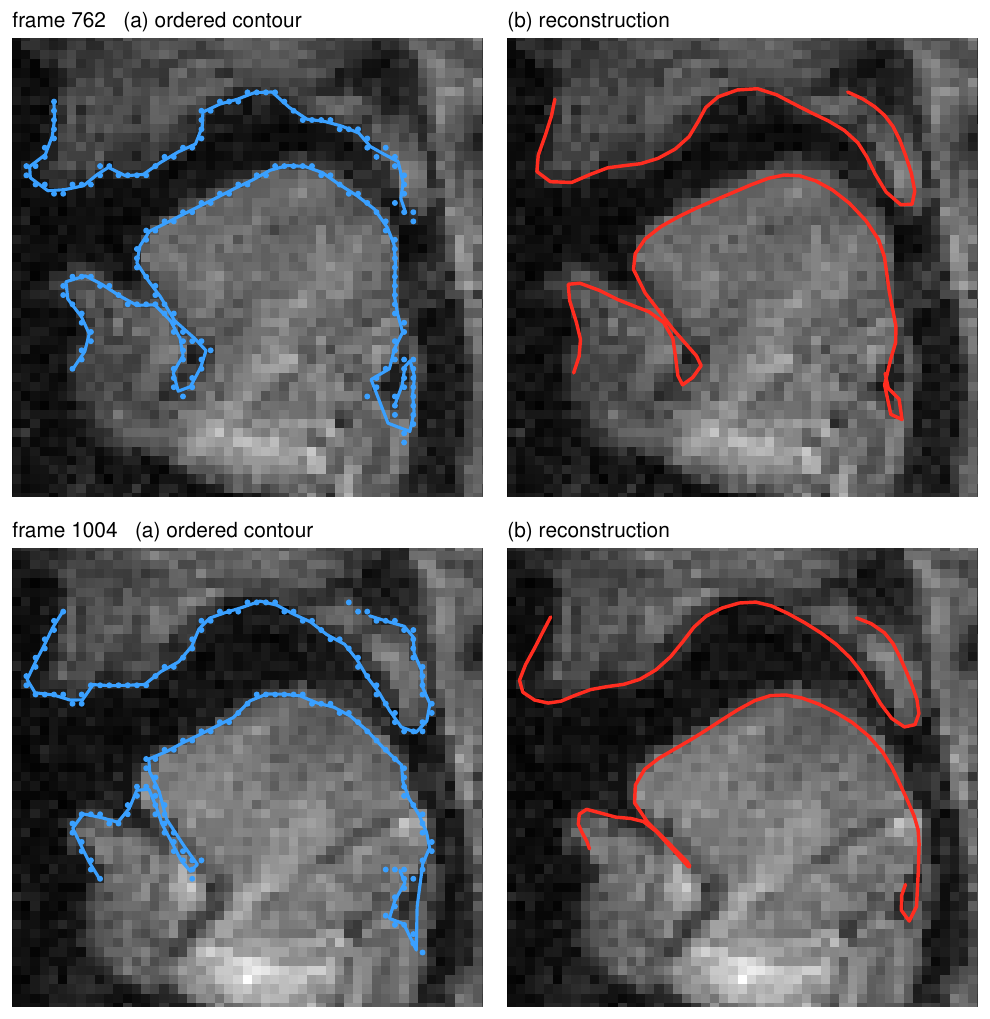}
 \caption{Two frames whose ordered contour does not recover the whole segmentation, and their reconstruction by the temporal smoothing model. In the left panels, the points of the ordered contour and the spline through them (blue): the shape that the pre-processing of Section~\ref{sec:contour_ordering} produces for that frame. In the right panels, the shape predicted at the same time point by \ourMethod\ with the frame acquisition time as covariate (red). Both frames are among those flagged as unusable and are therefore absent from the training data, so the predicted shapes are reconstructed from the neighboring frames alone. Top row, frame $762$: the ordered curve shows a partial velum, whereas the fitted shape restores a full velum. Bottom row, frame $1004$: the network labels misclassified the lower lip, the fitted shape recovers a regular lower lip.\label{fig:frames_example}}
\end{figure}

Both competitors are linear in the covariate, having been designed for covariates such as a group indicator or age, and cannot follow the strongly nonlinear articulatory motion of a fourteen-second recording, which is why they remain at the level of the baseline. Therefore, smoothing a shape trajectory calls for a nonparametric link and our model fills this gap. 
A video of the whole recording is provided as supplementary material: it
shows, frame by frame and on the rt-MRI image, the ordered contour produced
by the pre-processing next to the shape reconstructed by the model, in the
layout of Figure~\ref{fig:frames_example}, which reports two of its frames. The video shows that, in frames where the tongue touches the palate, the reconstructed boundaries may intersect, as do the original segmented contours in approximately $3\%$ of frames. This occurs because the model does not impose physical constraints on the predicted shapes, however, these shapes do not correspond to physically plausible vocal-tract configurations. Incorporating a penalty for self-intersection would therefore be a natural extension. We leave this for future work, since the issue arises in only a very small proportion of frames.

\begin{figure}[t]
\centering
\begin{tikzpicture}[x=1pt,y=1pt]
\definecolor{fillColor}{RGB}{255,255,255}
\path[use as bounding box,fill=fillColor,fill opacity=0.00] (0,0) rectangle (224.04,209.58);
\begin{scope}
\path[clip] ( 38.88, 23.76) rectangle (218.64,187.98);
\definecolor{fillColor}{RGB}{169,169,169}

\path[fill=fillColor] ( 49.70, 60.09) --
	( 82.99, 60.09) --
	( 82.99, 91.14) --
	( 49.70, 91.14) --
	cycle;
\definecolor{drawColor}{RGB}{0,0,0}

\path[draw=drawColor,line width= 1.2pt,line join=round] ( 49.70, 74.22) -- ( 82.99, 74.22);

\path[draw=drawColor,line width= 0.4pt,dash pattern=on 4pt off 4pt ,line join=round,line cap=round] ( 66.34, 29.84) -- ( 66.34, 60.09);

\path[draw=drawColor,line width= 0.4pt,dash pattern=on 4pt off 4pt ,line join=round,line cap=round] ( 66.34,123.76) -- ( 66.34, 91.14);

\path[draw=drawColor,line width= 0.4pt,line join=round,line cap=round] ( 58.02, 29.84) -- ( 74.66, 29.84);

\path[draw=drawColor,line width= 0.4pt,line join=round,line cap=round] ( 58.02,123.76) -- ( 74.66,123.76);

\path[draw=drawColor,line width= 0.4pt,line join=round,line cap=round] ( 49.70, 60.09) --
	( 82.99, 60.09) --
	( 82.99, 91.14) --
	( 49.70, 91.14) --
	cycle;

\path[draw=drawColor,line width= 0.4pt,line join=round,line cap=round] ( 66.34,155.71) circle (  1.82);

\path[draw=drawColor,line width= 0.4pt,line join=round,line cap=round] ( 66.34,142.13) circle (  1.82);

\path[draw=drawColor,line width= 0.4pt,line join=round,line cap=round] ( 66.34,130.46) circle (  1.82);

\path[draw=drawColor,line width= 0.4pt,line join=round,line cap=round] ( 66.34,134.03) circle (  1.82);
\definecolor{fillColor}{RGB}{211,211,211}

\path[fill=fillColor] ( 91.31,116.60) --
	(124.60,116.60) --
	(124.60,148.31) --
	( 91.31,148.31) --
	cycle;

\path[draw=drawColor,line width= 1.2pt,line join=round] ( 91.31,134.69) -- (124.60,134.69);

\path[draw=drawColor,line width= 0.4pt,dash pattern=on 4pt off 4pt ,line join=round,line cap=round] (107.95, 99.65) -- (107.95,116.60);

\path[draw=drawColor,line width= 0.4pt,dash pattern=on 4pt off 4pt ,line join=round,line cap=round] (107.95,181.90) -- (107.95,148.31);

\path[draw=drawColor,line width= 0.4pt,line join=round,line cap=round] ( 99.63, 99.65) -- (116.28, 99.65);

\path[draw=drawColor,line width= 0.4pt,line join=round,line cap=round] ( 99.63,181.90) -- (116.28,181.90);

\path[draw=drawColor,line width= 0.4pt,line join=round,line cap=round] ( 91.31,116.60) --
	(124.60,116.60) --
	(124.60,148.31) --
	( 91.31,148.31) --
	cycle;

\path[fill=fillColor] (132.92, 90.92) --
	(166.21, 90.92) --
	(166.21,108.62) --
	(132.92,108.62) --
	cycle;

\path[draw=drawColor,line width= 1.2pt,line join=round] (132.92, 97.02) -- (166.21, 97.02);

\path[draw=drawColor,line width= 0.4pt,dash pattern=on 4pt off 4pt ,line join=round,line cap=round] (149.56, 69.88) -- (149.56, 90.92);

\path[draw=drawColor,line width= 0.4pt,dash pattern=on 4pt off 4pt ,line join=round,line cap=round] (149.56,127.79) -- (149.56,108.62);

\path[draw=drawColor,line width= 0.4pt,line join=round,line cap=round] (141.24, 69.88) -- (157.89, 69.88);

\path[draw=drawColor,line width= 0.4pt,line join=round,line cap=round] (141.24,127.79) -- (157.89,127.79);

\path[draw=drawColor,line width= 0.4pt,line join=round,line cap=round] (132.92, 90.92) --
	(166.21, 90.92) --
	(166.21,108.62) --
	(132.92,108.62) --
	cycle;

\path[draw=drawColor,line width= 0.4pt,line join=round,line cap=round] (149.56,160.33) circle (  1.82);

\path[draw=drawColor,line width= 0.4pt,line join=round,line cap=round] (149.56,157.95) circle (  1.82);

\path[draw=drawColor,line width= 0.4pt,line join=round,line cap=round] (149.56,138.46) circle (  1.82);

\path[draw=drawColor,line width= 0.4pt,line join=round,line cap=round] (149.56,141.70) circle (  1.82);

\path[draw=drawColor,line width= 0.4pt,line join=round,line cap=round] (149.56,135.53) circle (  1.82);

\path[fill=fillColor] (174.53, 91.17) --
	(207.82, 91.17) --
	(207.82,108.71) --
	(174.53,108.71) --
	cycle;

\path[draw=drawColor,line width= 1.2pt,line join=round] (174.53, 97.47) -- (207.82, 97.47);

\path[draw=drawColor,line width= 0.4pt,dash pattern=on 4pt off 4pt ,line join=round,line cap=round] (191.17, 70.29) -- (191.17, 91.17);

\path[draw=drawColor,line width= 0.4pt,dash pattern=on 4pt off 4pt ,line join=round,line cap=round] (191.17,127.48) -- (191.17,108.71);

\path[draw=drawColor,line width= 0.4pt,line join=round,line cap=round] (182.85, 70.29) -- (199.50, 70.29);

\path[draw=drawColor,line width= 0.4pt,line join=round,line cap=round] (182.85,127.48) -- (199.50,127.48);

\path[draw=drawColor,line width= 0.4pt,line join=round,line cap=round] (174.53, 91.17) --
	(207.82, 91.17) --
	(207.82,108.71) --
	(174.53,108.71) --
	cycle;

\path[draw=drawColor,line width= 0.4pt,line join=round,line cap=round] (191.17,161.03) circle (  1.82);

\path[draw=drawColor,line width= 0.4pt,line join=round,line cap=round] (191.17,158.08) circle (  1.82);

\path[draw=drawColor,line width= 0.4pt,line join=round,line cap=round] (191.17,137.80) circle (  1.82);

\path[draw=drawColor,line width= 0.4pt,line join=round,line cap=round] (191.17,143.04) circle (  1.82);

\path[draw=drawColor,line width= 0.4pt,line join=round,line cap=round] (191.17,138.46) circle (  1.82);

\path[draw=drawColor,line width= 0.4pt,line join=round,line cap=round] (191.17,132.47) circle (  1.82);
\end{scope}
\begin{scope}
\path[clip] (  0.00,  0.00) rectangle (224.04,209.58);
\definecolor{drawColor}{RGB}{0,0,0}

\path[draw=drawColor,line width= 0.4pt,line join=round,line cap=round] ( 66.34, 23.76) -- (191.17, 23.76);

\path[draw=drawColor,line width= 0.4pt,line join=round,line cap=round] ( 66.34, 23.76) -- ( 66.34, 18.36);

\path[draw=drawColor,line width= 0.4pt,line join=round,line cap=round] (107.95, 23.76) -- (107.95, 18.36);

\path[draw=drawColor,line width= 0.4pt,line join=round,line cap=round] (149.56, 23.76) -- (149.56, 18.36);

\path[draw=drawColor,line width= 0.4pt,line join=round,line cap=round] (191.17, 23.76) -- (191.17, 18.36);

\node[text=drawColor,anchor=base,inner sep=0pt, outer sep=0pt, scale=  0.90] at ( 66.34,  8.64) {\texttt{ESR}};

\node[text=drawColor,anchor=base,inner sep=0pt, outer sep=0pt, scale=  0.90] at (107.95,  8.64) {\texttt{STEYER}};

\node[text=drawColor,anchor=base,inner sep=0pt, outer sep=0pt, scale=  0.90] at (149.56,  8.64) {\texttt{GUO}};

\node[text=drawColor,anchor=base,inner sep=0pt, outer sep=0pt, scale=  0.90] at (191.17,  8.64) {\texttt{FM}};

\path[draw=drawColor,line width= 0.4pt,line join=round,line cap=round] ( 38.88, 85.25) -- ( 38.88,175.52);

\path[draw=drawColor,line width= 0.4pt,line join=round,line cap=round] ( 38.88, 85.25) -- ( 33.48, 85.25);

\path[draw=drawColor,line width= 0.4pt,line join=round,line cap=round] ( 38.88,125.19) -- ( 33.48,125.19);

\path[draw=drawColor,line width= 0.4pt,line join=round,line cap=round] ( 38.88,153.53) -- ( 33.48,153.53);

\path[draw=drawColor,line width= 0.4pt,line join=round,line cap=round] ( 38.88,175.52) -- ( 33.48,175.52);

\node[text=drawColor,anchor=base east,inner sep=0pt, outer sep=0pt, scale=  0.90] at ( 32.40, 82.15) {0.2};

\node[text=drawColor,anchor=base east,inner sep=0pt, outer sep=0pt, scale=  0.90] at ( 32.40,122.10) {0.3};

\node[text=drawColor,anchor=base east,inner sep=0pt, outer sep=0pt, scale=  0.90] at ( 32.40,150.43) {0.4};

\node[text=drawColor,anchor=base east,inner sep=0pt, outer sep=0pt, scale=  0.90] at ( 32.40,172.42) {0.5};
\end{scope}
\begin{scope}
\path[clip] (  0.00,  0.00) rectangle (224.04,209.58);
\definecolor{drawColor}{RGB}{0,0,0}

\node[text=drawColor,anchor=base,inner sep=0pt, outer sep=0pt, scale=  1.08] at (128.76,195.06) {Missing-frame recovery};

\node[text=drawColor,rotate= 90.00,anchor=base,inner sep=0pt, outer sep=0pt, scale=  0.90] at ( 10.80,105.87) {elastic distance (log scale)};
\end{scope}
\begin{scope}
\path[clip] (  0.00,  0.00) rectangle (224.04,209.58);
\definecolor{drawColor}{RGB}{0,0,0}

\path[draw=drawColor,line width= 0.4pt,line join=round,line cap=round] ( 38.88, 23.76) --
	(218.64, 23.76) --
	(218.64,187.98) --
	( 38.88,187.98) --
	cycle;
\end{scope}
\end{tikzpicture}
\begin{tikzpicture}[x=1pt,y=1pt]
\definecolor{fillColor}{RGB}{255,255,255}
\path[use as bounding box,fill=fillColor,fill opacity=0.00] (0,0) rectangle (224.04,209.58);
\begin{scope}
\path[clip] ( 38.88, 23.76) rectangle (218.64,187.98);
\definecolor{fillColor}{RGB}{169,169,169}

\path[fill=fillColor] ( 49.70, 63.56) --
	( 82.99, 63.56) --
	( 82.99, 92.37) --
	( 49.70, 92.37) --
	cycle;
\definecolor{drawColor}{RGB}{0,0,0}

\path[draw=drawColor,line width= 1.2pt,line join=round] ( 49.70, 75.88) -- ( 82.99, 75.88);

\path[draw=drawColor,line width= 0.4pt,dash pattern=on 4pt off 4pt ,line join=round,line cap=round] ( 66.34, 29.84) -- ( 66.34, 63.56);

\path[draw=drawColor,line width= 0.4pt,dash pattern=on 4pt off 4pt ,line join=round,line cap=round] ( 66.34,123.45) -- ( 66.34, 92.37);

\path[draw=drawColor,line width= 0.4pt,line join=round,line cap=round] ( 58.02, 29.84) -- ( 74.66, 29.84);

\path[draw=drawColor,line width= 0.4pt,line join=round,line cap=round] ( 58.02,123.45) -- ( 74.66,123.45);

\path[draw=drawColor,line width= 0.4pt,line join=round,line cap=round] ( 49.70, 63.56) --
	( 82.99, 63.56) --
	( 82.99, 92.37) --
	( 49.70, 92.37) --
	cycle;

\path[draw=drawColor,line width= 0.4pt,line join=round,line cap=round] ( 66.34,127.82) circle (  1.82);

\path[draw=drawColor,line width= 0.4pt,line join=round,line cap=round] ( 66.34,129.33) circle (  1.82);

\path[draw=drawColor,line width= 0.4pt,line join=round,line cap=round] ( 66.34,125.48) circle (  1.82);

\path[draw=drawColor,line width= 0.4pt,line join=round,line cap=round] ( 66.34,160.81) circle (  1.82);

\path[draw=drawColor,line width= 0.4pt,line join=round,line cap=round] ( 66.34,139.22) circle (  1.82);

\path[draw=drawColor,line width= 0.4pt,line join=round,line cap=round] ( 66.34,133.74) circle (  1.82);

\path[draw=drawColor,line width= 0.4pt,line join=round,line cap=round] ( 66.34,125.41) circle (  1.82);

\path[draw=drawColor,line width= 0.4pt,line join=round,line cap=round] ( 66.34,126.78) circle (  1.82);

\path[draw=drawColor,line width= 0.4pt,line join=round,line cap=round] ( 66.34,124.84) circle (  1.82);
\definecolor{fillColor}{RGB}{211,211,211}

\path[fill=fillColor] ( 91.31,119.78) --
	(124.60,119.78) --
	(124.60,147.47) --
	( 91.31,147.47) --
	cycle;

\path[draw=drawColor,line width= 1.2pt,line join=round] ( 91.31,135.93) -- (124.60,135.93);

\path[draw=drawColor,line width= 0.4pt,dash pattern=on 4pt off 4pt ,line join=round,line cap=round] (107.95, 98.75) -- (107.95,119.78);

\path[draw=drawColor,line width= 0.4pt,dash pattern=on 4pt off 4pt ,line join=round,line cap=round] (107.95,164.64) -- (107.95,147.47);

\path[draw=drawColor,line width= 0.4pt,line join=round,line cap=round] ( 99.63, 98.75) -- (116.28, 98.75);

\path[draw=drawColor,line width= 0.4pt,line join=round,line cap=round] ( 99.63,164.64) -- (116.28,164.64);

\path[draw=drawColor,line width= 0.4pt,line join=round,line cap=round] ( 91.31,119.78) --
	(124.60,119.78) --
	(124.60,147.47) --
	( 91.31,147.47) --
	cycle;

\path[draw=drawColor,line width= 0.4pt,line join=round,line cap=round] (107.95,181.90) circle (  1.82);

\path[fill=fillColor] (132.92, 89.48) --
	(166.21, 89.48) --
	(166.21,107.30) --
	(132.92,107.30) --
	cycle;

\path[draw=drawColor,line width= 1.2pt,line join=round] (132.92, 97.29) -- (166.21, 97.29);

\path[draw=drawColor,line width= 0.4pt,dash pattern=on 4pt off 4pt ,line join=round,line cap=round] (149.56, 72.05) -- (149.56, 89.48);

\path[draw=drawColor,line width= 0.4pt,dash pattern=on 4pt off 4pt ,line join=round,line cap=round] (149.56,128.27) -- (149.56,107.30);

\path[draw=drawColor,line width= 0.4pt,line join=round,line cap=round] (141.24, 72.05) -- (157.89, 72.05);

\path[draw=drawColor,line width= 0.4pt,line join=round,line cap=round] (141.24,128.27) -- (157.89,128.27);

\path[draw=drawColor,line width= 0.4pt,line join=round,line cap=round] (132.92, 89.48) --
	(166.21, 89.48) --
	(166.21,107.30) --
	(132.92,107.30) --
	cycle;

\path[draw=drawColor,line width= 0.4pt,line join=round,line cap=round] (149.56,136.24) circle (  1.82);

\path[draw=drawColor,line width= 0.4pt,line join=round,line cap=round] (149.56,132.43) circle (  1.82);

\path[draw=drawColor,line width= 0.4pt,line join=round,line cap=round] (149.56,172.84) circle (  1.82);

\path[draw=drawColor,line width= 0.4pt,line join=round,line cap=round] (149.56,160.10) circle (  1.82);

\path[draw=drawColor,line width= 0.4pt,line join=round,line cap=round] (149.56,133.68) circle (  1.82);

\path[draw=drawColor,line width= 0.4pt,line join=round,line cap=round] (149.56,129.34) circle (  1.82);

\path[draw=drawColor,line width= 0.4pt,line join=round,line cap=round] (149.56,130.65) circle (  1.82);

\path[draw=drawColor,line width= 0.4pt,line join=round,line cap=round] (149.56,145.35) circle (  1.82);

\path[draw=drawColor,line width= 0.4pt,line join=round,line cap=round] (149.56,133.80) circle (  1.82);

\path[draw=drawColor,line width= 0.4pt,line join=round,line cap=round] (149.56,139.61) circle (  1.82);

\path[fill=fillColor] (174.53, 88.62) --
	(207.82, 88.62) --
	(207.82,107.49) --
	(174.53,107.49) --
	cycle;

\path[draw=drawColor,line width= 1.2pt,line join=round] (174.53, 97.59) -- (207.82, 97.59);

\path[draw=drawColor,line width= 0.4pt,dash pattern=on 4pt off 4pt ,line join=round,line cap=round] (191.17, 72.03) -- (191.17, 88.62);

\path[draw=drawColor,line width= 0.4pt,dash pattern=on 4pt off 4pt ,line join=round,line cap=round] (191.17,129.20) -- (191.17,107.49);

\path[draw=drawColor,line width= 0.4pt,line join=round,line cap=round] (182.85, 72.03) -- (199.50, 72.03);

\path[draw=drawColor,line width= 0.4pt,line join=round,line cap=round] (182.85,129.20) -- (199.50,129.20);

\path[draw=drawColor,line width= 0.4pt,line join=round,line cap=round] (174.53, 88.62) --
	(207.82, 88.62) --
	(207.82,107.49) --
	(174.53,107.49) --
	cycle;

\path[draw=drawColor,line width= 0.4pt,line join=round,line cap=round] (191.17,137.39) circle (  1.82);

\path[draw=drawColor,line width= 0.4pt,line join=round,line cap=round] (191.17,134.87) circle (  1.82);

\path[draw=drawColor,line width= 0.4pt,line join=round,line cap=round] (191.17,157.26) circle (  1.82);

\path[draw=drawColor,line width= 0.4pt,line join=round,line cap=round] (191.17,159.78) circle (  1.82);

\path[draw=drawColor,line width= 0.4pt,line join=round,line cap=round] (191.17,133.89) circle (  1.82);

\path[draw=drawColor,line width= 0.4pt,line join=round,line cap=round] (191.17,132.44) circle (  1.82);

\path[draw=drawColor,line width= 0.4pt,line join=round,line cap=round] (191.17,132.34) circle (  1.82);

\path[draw=drawColor,line width= 0.4pt,line join=round,line cap=round] (191.17,143.93) circle (  1.82);

\path[draw=drawColor,line width= 0.4pt,line join=round,line cap=round] (191.17,135.60) circle (  1.82);

\path[draw=drawColor,line width= 0.4pt,line join=round,line cap=round] (191.17,137.11) circle (  1.82);
\end{scope}
\begin{scope}
\path[clip] (  0.00,  0.00) rectangle (224.04,209.58);
\definecolor{drawColor}{RGB}{0,0,0}

\path[draw=drawColor,line width= 0.4pt,line join=round,line cap=round] ( 66.34, 23.76) -- (191.17, 23.76);

\path[draw=drawColor,line width= 0.4pt,line join=round,line cap=round] ( 66.34, 23.76) -- ( 66.34, 18.36);

\path[draw=drawColor,line width= 0.4pt,line join=round,line cap=round] (107.95, 23.76) -- (107.95, 18.36);

\path[draw=drawColor,line width= 0.4pt,line join=round,line cap=round] (149.56, 23.76) -- (149.56, 18.36);

\path[draw=drawColor,line width= 0.4pt,line join=round,line cap=round] (191.17, 23.76) -- (191.17, 18.36);

\node[text=drawColor,anchor=base,inner sep=0pt, outer sep=0pt, scale=  0.90] at ( 66.34,  8.64) {\texttt{ESR}};

\node[text=drawColor,anchor=base,inner sep=0pt, outer sep=0pt, scale=  0.90] at (107.95,  8.64) {\texttt{STEYER}};

\node[text=drawColor,anchor=base,inner sep=0pt, outer sep=0pt, scale=  0.90] at (149.56,  8.64) {\texttt{GUO}};

\node[text=drawColor,anchor=base,inner sep=0pt, outer sep=0pt, scale=  0.90] at (191.17,  8.64) {\texttt{FM}};

\path[draw=drawColor,line width= 0.4pt,line join=round,line cap=round] ( 38.88, 84.96) -- ( 38.88,174.97);

\path[draw=drawColor,line width= 0.4pt,line join=round,line cap=round] ( 38.88, 84.96) -- ( 33.48, 84.96);

\path[draw=drawColor,line width= 0.4pt,line join=round,line cap=round] ( 38.88,124.79) -- ( 33.48,124.79);

\path[draw=drawColor,line width= 0.4pt,line join=round,line cap=round] ( 38.88,153.05) -- ( 33.48,153.05);

\path[draw=drawColor,line width= 0.4pt,line join=round,line cap=round] ( 38.88,174.97) -- ( 33.48,174.97);

\node[text=drawColor,anchor=base east,inner sep=0pt, outer sep=0pt, scale=  0.90] at ( 32.40, 81.86) {0.2};

\node[text=drawColor,anchor=base east,inner sep=0pt, outer sep=0pt, scale=  0.90] at ( 32.40,121.69) {0.3};

\node[text=drawColor,anchor=base east,inner sep=0pt, outer sep=0pt, scale=  0.90] at ( 32.40,149.95) {0.4};

\node[text=drawColor,anchor=base east,inner sep=0pt, outer sep=0pt, scale=  0.90] at ( 32.40,171.87) {0.5};
\end{scope}
\begin{scope}
\path[clip] (  0.00,  0.00) rectangle (224.04,209.58);
\definecolor{drawColor}{RGB}{0,0,0}

\node[text=drawColor,anchor=base,inner sep=0pt, outer sep=0pt, scale=  1.08] at (128.76,195.06) {Low frame rate};

\node[text=drawColor,rotate= 90.00,anchor=base,inner sep=0pt, outer sep=0pt, scale=  0.90] at ( 10.80,105.87) {elastic distance (log scale)};
\end{scope}
\begin{scope}
\path[clip] (  0.00,  0.00) rectangle (224.04,209.58);
\definecolor{drawColor}{RGB}{0,0,0}

\path[draw=drawColor,line width= 0.4pt,line join=round,line cap=round] ( 38.88, 23.76) --
	(218.64, 23.76) --
	(218.64,187.98) --
	( 38.88,187.98) --
	cycle;
\end{scope}
\end{tikzpicture}
 \caption{Temporal smoothing of the vocal tract trajectory: elastic distance
  $\shapeDist$ between each held-out shape and its prediction, lower being
  better. Boxes, from left to right: \ourMethod\ (dark grey), the quotient
  linear regression of \citet{steyerRegressionQuotientMetric2023} (\texttt{STEYER}),
  the tangent-space regression on principal component scores of
  \citet{guo2020statistical} (\texttt{GUO}), and the Fréchet mean of the training
  shapes as covariate-free baseline (\texttt{FM}). Left, missing-frame recovery: one chunk of five consecutive
  frames removed from every window of twenty ($60$ ms gaps, $250$ held-out
  frames). Right, low frame rate: of every four frames the first is used for the fit and the
  third for the test, the other two discarded ($252$ held-out frames),
  mimicking an acquisition four times slower than the
  recording.\label{fig:boxplots_smoothing}}
\end{figure}

\section{Speech inversion}

\label{sec:speech inversion}

\subsection{Overview}

Speech inversion estimates time-varying articulatory configurations from acoustic
signals. The inverse mapping is nonlinear and ill-posed: distinct vocal-tract
shapes can produce similar acoustics, even in simplified tube models, and
compensatory articulation further increases this ambiguity
\citep{maeda1990compensatory,mrayati_distinctive_1988}.

Early methods used acoustic models, articulatory synthesis, or codebooks
\citep{schroeder1967determination,mermelstein1967determination,Tukey1978InversionOA},
whereas later work used X-ray microbeam (XRMB) or electromagnetic articulography (EMA) databases for statistical, dynamical-system,
and deep-learning inversion 
\citep[see][and references therein]{tavakoli2025statistics}.

Because XRMB and EMA use only a handful of sensors, recent studies
have instead predicted midsagittal rt-MRI images or articulator contours from
acoustics \citep{csapó2020speakerAAI,li2015estimate,nguyen2024speech2rtmri,azzouz2025complete,
azzouz2025reconstruction,azzouz2026embeddings}. We refer to \citet{tavakoli2025statistics} for a
broader review. We showcase our method for speech inversion based on rt-MRI data.

\subsection{Model fitting and results}

We use the recording introduced in Section~\ref{sec:smoothing} and take only the
tongue contour as the response. The preprocessing follows Section~\ref{sec:contour_ordering}.
This yields usable contours for $1109$ of the $1115$ segmented
frames. The acoustic predictors are the first $20$ principal components of
MFCCs together with their first and second time differences. Details of the
acoustic preprocessing and hyperparameters are given in the Appendix. The
nine missing or unusable training contours are completed using the temporal
smoothing fit from Section~\ref{sec:smoothing}.

The recording has four sections, each repeating /a/, /i/, and /u/ around a
different consonant: /p/, /t/, /k/, and /m/. We hold out the first section
(frames $41$--$284$) and fit on the remaining three, leaving $240$ test frames
and $843$ training frames ($834$ extracted and $9$ completed) after removing
silent pauses. Importantly, the test section is separated from training in time and contains
the \emph{unseen} consonant /p/. We fit \ourMethod\ with a GRBF kernel, selecting its
bandwidth and regularization parameter by four-fold cross-validation on
contiguous training blocks. We compare it with the Fréchet mean of the
training shapes and evaluate predictions using the elastic distance
$\shapeDist$.

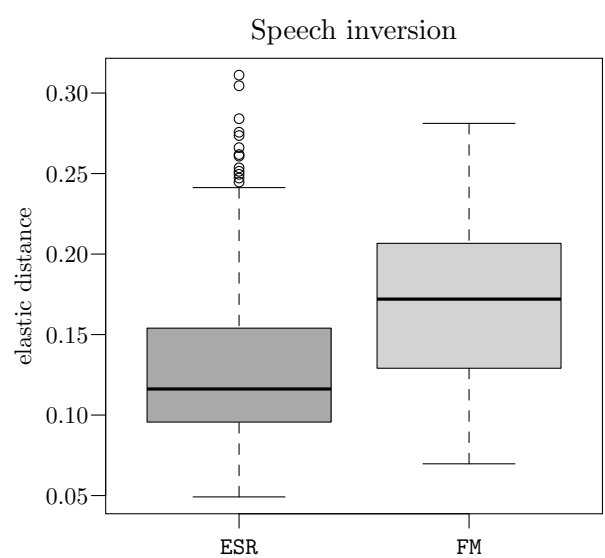
\begin{figure}[t]
\centering
\begin{tikzpicture}[x=1pt,y=1pt]
\definecolor{fillColor}{RGB}{255,255,255}
\path[use as bounding box,fill=fillColor,fill opacity=0.00] (0,0) rectangle (231.26,216.81);
\begin{scope}
\path[clip] ( 38.88, 23.76) rectangle (225.86,195.21);
\definecolor{fillColor}{RGB}{169,169,169}

\path[fill=fillColor] ( 54.46, 58.26) --
	(123.72, 58.26) --
	(123.72, 93.63) --
	( 54.46, 93.63) --
	cycle;
\definecolor{drawColor}{RGB}{0,0,0}

\path[draw=drawColor,line width= 1.2pt,line join=round] ( 54.46, 70.73) -- (123.72, 70.73);

\path[draw=drawColor,line width= 0.4pt,dash pattern=on 4pt off 4pt ,line join=round,line cap=round] ( 89.09, 30.11) -- ( 89.09, 58.26);

\path[draw=drawColor,line width= 0.4pt,dash pattern=on 4pt off 4pt ,line join=round,line cap=round] ( 89.09,146.53) -- ( 89.09, 93.63);

\path[draw=drawColor,line width= 0.4pt,line join=round,line cap=round] ( 71.78, 30.11) -- (106.40, 30.11);

\path[draw=drawColor,line width= 0.4pt,line join=round,line cap=round] ( 71.78,146.53) -- (106.40,146.53);

\path[draw=drawColor,line width= 0.4pt,line join=round,line cap=round] ( 54.46, 58.26) --
	(123.72, 58.26) --
	(123.72, 93.63) --
	( 54.46, 93.63) --
	cycle;

\path[draw=drawColor,line width= 0.4pt,line join=round,line cap=round] ( 89.09,184.85) circle (  1.82);

\path[draw=drawColor,line width= 0.4pt,line join=round,line cap=round] ( 89.09,150.12) circle (  1.82);

\path[draw=drawColor,line width= 0.4pt,line join=round,line cap=round] ( 89.09,148.62) circle (  1.82);

\path[draw=drawColor,line width= 0.4pt,line join=round,line cap=round] ( 89.09,167.37) circle (  1.82);

\path[draw=drawColor,line width= 0.4pt,line join=round,line cap=round] ( 89.09,151.52) circle (  1.82);

\path[draw=drawColor,line width= 0.4pt,line join=round,line cap=round] ( 89.09,158.38) circle (  1.82);

\path[draw=drawColor,line width= 0.4pt,line join=round,line cap=round] ( 89.09,153.97) circle (  1.82);

\path[draw=drawColor,line width= 0.4pt,line join=round,line cap=round] ( 89.09,158.97) circle (  1.82);

\path[draw=drawColor,line width= 0.4pt,line join=round,line cap=round] ( 89.09,152.67) circle (  1.82);

\path[draw=drawColor,line width= 0.4pt,line join=round,line cap=round] ( 89.09,166.12) circle (  1.82);

\path[draw=drawColor,line width= 0.4pt,line join=round,line cap=round] ( 89.09,161.57) circle (  1.82);

\path[draw=drawColor,line width= 0.4pt,line join=round,line cap=round] ( 89.09,172.45) circle (  1.82);

\path[draw=drawColor,line width= 0.4pt,line join=round,line cap=round] ( 89.09,188.86) circle (  1.82);
\definecolor{fillColor}{RGB}{211,211,211}

\path[fill=fillColor] (141.03, 78.53) --
	(210.28, 78.53) --
	(210.28,125.55) --
	(141.03,125.55) --
	cycle;

\path[draw=drawColor,line width= 1.2pt,line join=round] (141.03,104.55) -- (210.28,104.55);

\path[draw=drawColor,line width= 0.4pt,dash pattern=on 4pt off 4pt ,line join=round,line cap=round] (175.66, 42.56) -- (175.66, 78.53);

\path[draw=drawColor,line width= 0.4pt,dash pattern=on 4pt off 4pt ,line join=round,line cap=round] (175.66,170.70) -- (175.66,125.55);

\path[draw=drawColor,line width= 0.4pt,line join=round,line cap=round] (158.34, 42.56) -- (192.97, 42.56);

\path[draw=drawColor,line width= 0.4pt,line join=round,line cap=round] (158.34,170.70) -- (192.97,170.70);

\path[draw=drawColor,line width= 0.4pt,line join=round,line cap=round] (141.03, 78.53) --
	(210.28, 78.53) --
	(210.28,125.55) --
	(141.03,125.55) --
	cycle;
\end{scope}
\begin{scope}
\path[clip] (  0.00,  0.00) rectangle (231.26,216.81);
\definecolor{drawColor}{RGB}{0,0,0}

\path[draw=drawColor,line width= 0.4pt,line join=round,line cap=round] ( 89.09, 23.76) -- (175.66, 23.76);

\path[draw=drawColor,line width= 0.4pt,line join=round,line cap=round] ( 89.09, 23.76) -- ( 89.09, 18.36);

\path[draw=drawColor,line width= 0.4pt,line join=round,line cap=round] (175.66, 23.76) -- (175.66, 18.36);

\node[text=drawColor,anchor=base,inner sep=0pt, outer sep=0pt, scale=  0.90] at ( 89.09,  8.64) {\texttt{ESR}};

\node[text=drawColor,anchor=base,inner sep=0pt, outer sep=0pt, scale=  0.90] at (175.66,  8.64) {\texttt{FM}};

\path[draw=drawColor,line width= 0.4pt,line join=round,line cap=round] ( 38.88, 30.61) -- ( 38.88,182.12);

\path[draw=drawColor,line width= 0.4pt,line join=round,line cap=round] ( 38.88, 30.61) -- ( 33.48, 30.61);

\path[draw=drawColor,line width= 0.4pt,line join=round,line cap=round] ( 38.88, 60.91) -- ( 33.48, 60.91);

\path[draw=drawColor,line width= 0.4pt,line join=round,line cap=round] ( 38.88, 91.21) -- ( 33.48, 91.21);

\path[draw=drawColor,line width= 0.4pt,line join=round,line cap=round] ( 38.88,121.51) -- ( 33.48,121.51);

\path[draw=drawColor,line width= 0.4pt,line join=round,line cap=round] ( 38.88,151.81) -- ( 33.48,151.81);

\path[draw=drawColor,line width= 0.4pt,line join=round,line cap=round] ( 38.88,182.12) -- ( 33.48,182.12);

\node[text=drawColor,anchor=base east,inner sep=0pt, outer sep=0pt, scale=  0.90] at ( 32.40, 27.51) {0.05};

\node[text=drawColor,anchor=base east,inner sep=0pt, outer sep=0pt, scale=  0.90] at ( 32.40, 57.81) {0.10};

\node[text=drawColor,anchor=base east,inner sep=0pt, outer sep=0pt, scale=  0.90] at ( 32.40, 88.11) {0.15};

\node[text=drawColor,anchor=base east,inner sep=0pt, outer sep=0pt, scale=  0.90] at ( 32.40,118.41) {0.20};

\node[text=drawColor,anchor=base east,inner sep=0pt, outer sep=0pt, scale=  0.90] at ( 32.40,148.71) {0.25};

\node[text=drawColor,anchor=base east,inner sep=0pt, outer sep=0pt, scale=  0.90] at ( 32.40,179.02) {0.30};
\end{scope}
\begin{scope}
\path[clip] (  0.00,  0.00) rectangle (231.26,216.81);
\definecolor{drawColor}{RGB}{0,0,0}

\node[text=drawColor,anchor=base,inner sep=0pt, outer sep=0pt, scale=  1.08] at (132.37,202.29) {Speech inversion};

\node[text=drawColor,rotate= 90.00,anchor=base,inner sep=0pt, outer sep=0pt, scale=  0.90] at ( 10.80,109.48) {elastic distance};
\end{scope}
\begin{scope}
\path[clip] (  0.00,  0.00) rectangle (231.26,216.81);
\definecolor{drawColor}{RGB}{0,0,0}

\path[draw=drawColor,line width= 0.4pt,line join=round,line cap=round] ( 38.88, 23.76) --
	(225.86, 23.76) --
	(225.86,195.21) --
	( 38.88,195.21) --
	cycle;
\end{scope}
\end{tikzpicture}
  \caption{Speech inversion: elastic distance between the held-out tongue
  shapes and the predictions of
  \ourMethod\ and of the Fréchet mean of the training
  shapes.\label{fig:boxplot_si}}
\end{figure}

\begin{figure}[t]
\centering
  \includegraphics[width=.7\textwidth]{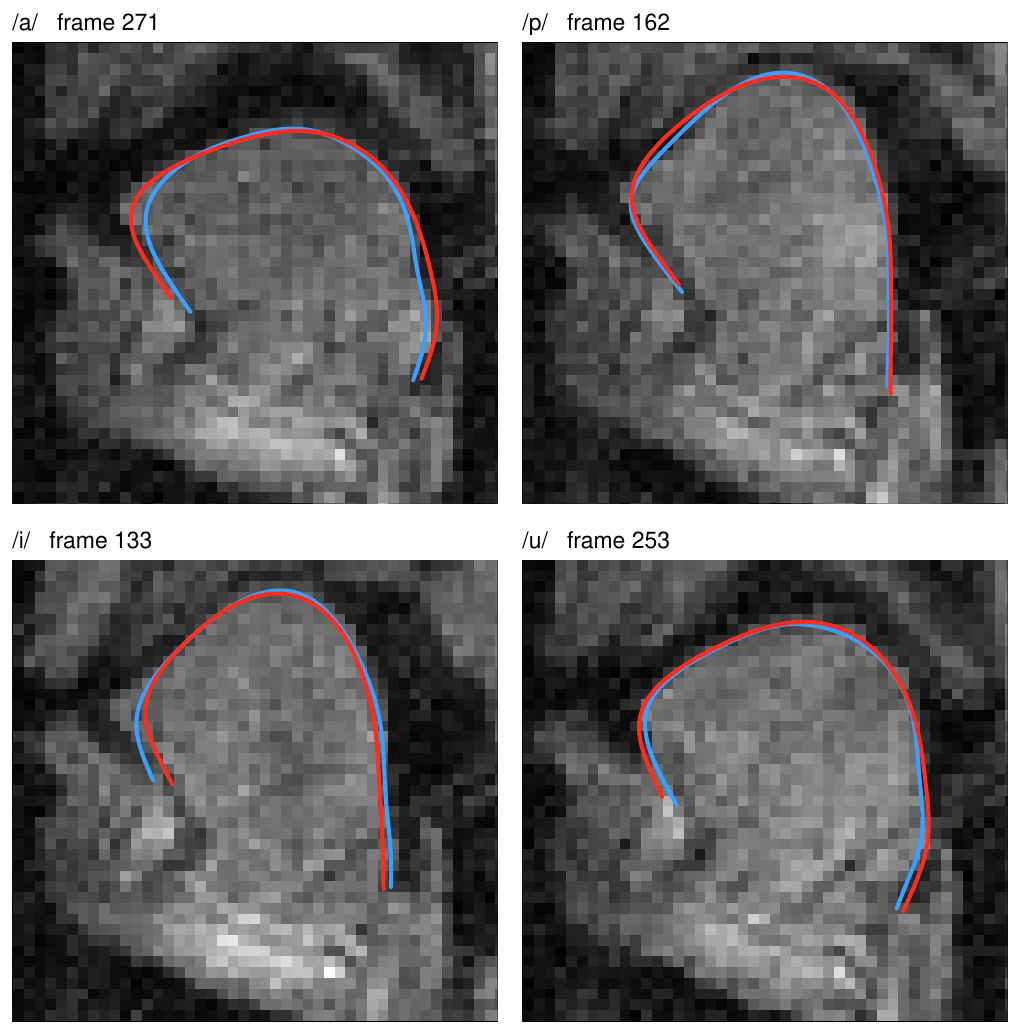}
  \caption{Speech inversion on four held-out frames, one for each phoneme of the held-out section: the tongue shape predicted from the acoustics (red) and the shape extracted from the segmentation of the same frame (blue), drawn on the rt-MRI image. Top left, /a/; top right, /p/; bottom left, /i/; bottom right, /u/. \label{fig:frames_si}}
\end{figure}

As shown in Figure~\ref{fig:boxplot_si}, \ourMethod\ attains a mean elastic
distance of $0.130$, compared with $0.170$ for the Fréchet mean, a $24\%$
improvement. It is closer to the held-out shape on $83\%$ of test frames, and
Figure~\ref{fig:frames_si} illustrates representative predictions. The
error remains, however, of the same order as the spread of the shapes around
their Fr\'echet mean.
To relate this result to what is, to our knowledge, the only other approach that predicts complete
tongue contours from acoustics \citep{azzouz2025complete,azzouz2026embeddings},
we also place each predicted shape on the observed tongue of its frame and
measure the root mean square error (RMSE) between the two contours in the image
(see details and discussion in the Appendix). Averaged over the test frames our RMSE is $1.9$ mm. \citet{azzouz2025complete,azzouz2026embeddings} report an RMSE of $2.3$ mm for a training dataset of $3.5$ hours (audio/rtMRI) and an RMSE of $2.8$ mm for a training dataset of $10$ minutes. 
The two settings are, however, not directly comparable as \citet{azzouz2025complete,azzouz2026embeddings}'s model must also predict the position and size of the
tongue in the image, whereas the SRVF representation separates these from
the shape. Our model concentrates on the latter and the position, size and rotation alignment are
computed separately. 
Importantly, our training set comprises about ten seconds of speech, and
it is encouraging that such a short acoustic/articulatory dataset suffices for an error of the same order of magnitude. 
The improvement over the Fr\'echet mean shows that the acoustics carry information on the tongue shape that the model recovers. 

Our single-speaker application demonstrates the method's predictive potential; a comprehensive solution to speech inversion is left for future work.

\section{Discussion}
\label{sec:discussion}

We have proposed a nonparametric regression model for elastic planar shapes,
in which the conditional Fr\'echet mean is modeled through a vector-valued
RKHS function taking values in a tangent space of the SRVF sphere, and the
alignment of each observation over reparametrization and rotation is
estimated jointly with the regression function rather than fixed in
preprocessing. We make the alignment affordable at every iteration of the
fit by introducing a new quasi-Newton method that exploits the scale invariance of the
reparametrization cost to work without the sphere constraint. In the simulations,
comparisons with models that fix the alignment beforehand or that linearize
in a tangent space show that estimating the alignment within the regression
and remaining on the sphere both matter. In the temporal smoothing of vocal
tract contours, the model improves on the existing elastic shape regression
models, whose linear link cannot follow the articulatory motion.

We have demonstrated that the method proposed can be applied to the problem of speech inversion. The results are, however, preliminary, given the data available. Our method improves
on the Fr\'echet mean baseline, but the competing models of
Section~\ref{sec:sims} performed comparably. This is to be
expected when one considers the amount of data: our recording provides
about ten seconds of training speech, whereas the only existing approaches
predicting complete articulator contours from acoustics are trained on up to
three and a half hours \citep{azzouz2025complete,azzouz2026embeddings}. 
While their corpus and code are not public, we give an indirect comparison in Section~\ref{sec:speech inversion}.
Several directions remain for future work. A reliable pipeline would first
require far more data, multi-speaker corpora would then call for speaker
effects, which the kernel formulation accommodates as covariates, and
phonetic information could be added alongside the acoustics
\citep{azzouz2026phonetic}. 

Although we have worked with planar curves throughout, the setting could be
restated for curves in $\Real^p$, $p \geq 2$, with no change in the theory. The SRVF
framework, the isometric action of rotations and reparametrizations, and
the closed-form optimal rotation carry over
\citep[Chapter~10]{srivastava2016functional}, and the spherical regression
is formulated on the unit sphere of $L^2([0,1],\Real^p)$ as it stands. An extension to higher dimensional curves requires only an adjustment of the implementation. 

Beyond
speech production, the method applies as is to shape trajectories indexed
by time or other covariates, such as organ contours in medical imaging
\citep{dogra2023confounder,steyerRegressionQuotientMetric2023,pfeuffer2026deep}
or bone and organism outlines in biology \citep{stocker2022flexible}.

Several theoretical questions remain open. The penalized risk is not convex
in the alignments and, even though the alternating algorithm does not
increase it at any step, it is not guaranteed that the iterates approach a
minimum. The alignment step itself relies on the existence of an optimal
reparametrization between two curves, which is still a delicate aspect in some cases
\citep{bruveris2016optimal}. Finally,
consistency and rates of convergence for estimators of the conditional
Fr\'echet mean in the elastic shape space, where the distance is itself the
value of an optimization problem, remain open problems.

\section*{Acknowledgements}

AS acknowledges funding from SNSF Grant 200020 207367.

The authors thank Krishna Somandepalli and Shri Narayanan for providing the rtMRI data and extracted contours.

The authors thank Anuj Srivastava, Xavier Pennec, Stephan Huckemann, and Victor Panaretos for helpful discussion.

Coding, figure preparation, and proofreading were assisted by Claude Code (Anthropic).

\appendix

\clearpage
\section*{APPENDIX}

\section{Curve reparametrization: related work, algorithms and a numerical comparison}
\label{app:reparametrization}

This appendix collects the material on the optimization over $\Gamma$ that Section 3.2 of the main paper summarizes: a review of the approaches proposed in the literature, the two schemes of \citet{huang2016riemannian} together with the one we propose, and a numerical comparison of all of them against the dynamic programming routines in current use.

\subsection{Related work}
\label{app:reparam_review}

The optimization over the reparametrization group is the computational bottleneck of elastic shape analysis, and a considerable variety of strategies has been proposed for it. Here, we focus on the methods for planar shapes. It is convenient to group them as follows.

\emph{Dynamic programming.} The standard approach, described in Section~\ref{app:reparam_dp}, restricts $\gamma$ to monotone piecewise-linear paths on a grid and finds the optimal such path recursively \citep[Appendix~B]{srivastava2016functional}. It is the method we compare against.

\emph{Riemannian optimization over the sphere of warpings.} Since $\dot\gamma > 0$, a reparametrization can be represented by $l = \sqrt{\dot\gamma}$, which lies on the unit sphere $\sphereL$ of $L^2([0,1],\mathbb{R})$, this representation goes back to the Fisher--Rao registration framework of \citet{srivastava2011registration}. \citet{huang2016riemannian} treat the alignment as an optimization on that manifold and propose a Riemannian steepest descent and a Riemannian limited-memory BFGS, implemented in the \texttt{C++} library \texttt{ROPTLIB} \citep{huang2018roptlib}. These are the schemes of Section~\ref{app:reparam_riemannian}, and the ones for which we provide an \texttt{R} implementation.

\emph{Optimization over the knots of a piecewise-linear model.} \citet{steyer2023elastic} model both curves as piecewise-linear SRV curves and optimize the positions of the knots of the warping directly, so that it is not tied to a fixed grid. The method is implemented in the \texttt{elasdics} package \citep{steyer2023elasdics} and supplies the alignment step of the quotient regression model of \citet{steyerRegressionQuotientMetric2023}, which we use as a competitor at the level of the regression model in the application of the main paper. We do not include it in the comparison of Section~\ref{app:reparam_simulation}: there the two curves share the same image by construction, so the constant-speed parametrizations from which the method starts already coincide and it would be initialized at the exact solution.

\emph{Exact algorithms for piecewise-linear curves.} For curves that are exactly piecewise linear, \citet{lahiri2015precise} characterize the structure of the optimal matching and give an algorithm that produces it exactly rather than approximately.

\emph{Bayesian and learning-based approaches.} Outside the deterministic-optimization setting, \citet{cheng2016bayesian} place Dirichlet priors on the warping and sample from the posterior, which quantifies the uncertainty in the alignment at a higher computational cost. More recently, \citet{celledoni2023deep} construct diffeomorphisms as compositions of elementary ones learned by a neural network. These are attractive for large batches of alignments but require a training stage, and we do not pursue them here.

\emph{Theory.} Finally, whether the infimum over $\Gamma$ is attained at all is a delicate question: \citet{bruveris2016optimal} shows that optimal reparametrizations exist for $C^1$ curves, and exhibits Lipschitz curves for which they do not. In our setting the curves are smoothing splines, so the former case applies.

\subsection{Curve reparametrization through Dynamic Programming}
\label{app:reparam_dp}
A standard method for curve reparametrization, optimizing over $\gamma$ while keeping $O$ fixed, is given by Dynamic Programming \citep[DP;][Appendix~B]{srivastava2016functional}. The DP method divides the square $[0,1]\times[0,1]$ into a grid of $N$ points and looks for the optimal piecewise-linear function that approximates a reparametrization $\gamma \in \Gamma$. The key requirement making DP applicable is that the cost function $H(\,\cdot\,, O)$ is additive over $[0,1]$. Then, viewing $\gamma$ as a monotone path from $(0,0)$ to $(1,1)$ restricted to the grid, the cost of a piecewise-linear path decomposes into a sum of segment costs, each depending only on the segment's endpoints, so that the optimal path can be computed recursively, node by node, across the grid. DP is accurate, but its cost is quadratic in the discretization of the curves, and since the alignment is the innermost step of every model fit, this is prohibitive for our application. Subsequent work has been devoted to making it cheaper \citep[e.g.][]{bernal2016fast}, but we do not compare against it here for lack of an \texttt{R} implementation. We compare our alignment strategy in Section~\ref{app:reparam_simulation} against the two DP routines of the \texttt{fdasrvf} package \citep{tucker2025fdasrvf}, the reference software for the square-root velocity framework. One (\texttt{DPQ}) searching on the sampling grid itself, and one that decouples the search grid from it and allows both to be non-uniform (\texttt{DPQ2}).

\subsection{Curve reparametrization through Riemannian optimization}
\label{app:reparam_riemannian}
Another approach is to formulate the reparametrization problem as an infinite-dimensional Riemannian optimization problem, as in \citet{huang2016riemannian}.
Since any $\gamma \in \Gamma$ is an increasing diffeomorphism of $[0,1]$, it satisfies
$
\gamma(0)=0,\quad \gamma(1)=1,\quad \dot\gamma(t)>0 \text{ a.e.\ on } [0,1].
$
The positivity of $\dot\gamma$ implies that there exists an $l \in L^2([0,1],\mathbb{R})$ such that $
\dot\gamma(t) = l^2(t)$ almost everywhere, and the boundary conditions give
$\int_0^1 l^2(t)\,dt = \gamma(1)-\gamma(0) = 1.$ Thus $l$ lies on the $L^2([0,1],\mathbb{R})$ unit sphere
$
\sphereL
  = \bigl\{\,l \in L^2([0,1],\mathbb{R}) : \int_0^1 l^2(t)\,dt = 1\,\bigr\},
$
and conversely any $l \in \sphereL$ defines a valid reparametrization via $\gamma(t) = \int_0^t l^2(s)\,ds$.
General Riemannian optimization techniques can be applied \citep{boumal2023intromanifolds} since $\sphereL$ is an infinite-dimensional Riemannian manifold.

A Riemannian Broyden--Fletcher--Goldfarb--Shanno (BFGS) scheme for this problem is described by \citet{huang2016riemannian} and implemented in their \texttt{C++} library \texttt{ROPTLIB} \citep{huang2018roptlib}. That code is not, however, reachable from \texttt{R} as the \texttt{ManifoldOptim} interface \citep{martin2020manifoldoptim} bundles the \texttt{ROPTLIB} sources but exposes only its solvers and part of its manifolds, and the sphere of $L^2([0,1],\mathbb{R})$ is not included. The \texttt{fdasrvf} package, the reference implementation of the square-root velocity toolbox in \texttt{R}, does expose a Riemannian BFGS option, but only for scalar functions. For planar curves in \texttt{R} the dynamic programming routines of Section~\ref{app:reparam_dp} are therefore in practice the only alternative available.

We therefore provide, in the open-source \texttt{sphereg} package \citep{sphereg}, what is to our knowledge the first \texttt{R} implementation of the Riemannian optimization over $\Gamma$ in the representation of \citet{huang2016riemannian}. Both of the schemes they consider are available, embedded in the block-coordinate relaxation over $\Gamma \times SO(2)$ of Section~3.2 of the main paper: the first-order Riemannian steepest descent of Algorithm~\ref{code:steepdesc}, whose step size is selected by the automatic Armijo backtracking line search of Algorithm~\ref{code:armijo}, following \citet[][Section~4.2]{absil2008optimization}, and a Riemannian limited-memory BFGS on $\sphereL$.

We also propose a new scheme, which is the default in our implementation and which the
experiments of Section~\ref{app:reparam_simulation} show to be the most efficient of the
three. It rests on an observation that removes the need for spherical constraint. For fixed $O \in SO(2)$, write
\begin{align*}
  \gamma_l(t) \;=\; \frac{\int_0^t l^2(s)\,ds}{\int_0^1 l^2(s)\,ds},
  \qquad
  F(l) \;:=\; H(\gamma_l, O),
  \qquad l \in L^2([0,1],\mathbb{R}) \setminus \{0\},
\end{align*}
so that $F$ extends to the whole punctured space the cost that Algorithm~\ref{code:steepdesc}
minimizes on $\sphereL$, and let $F|_{\sphereL}$ denote that restriction. Since
$\gamma_{c\,l} = \gamma_l$ for every $c > 0$, the map $F$ is positively homogeneous of degree
zero. Every ray $\{c\,l \mid c > 0\}$ therefore meets $\sphereL$ exactly once, at
$\hat l := l / \|l\|_{L^2}$, and $F$ is constant along it, so that
$\inf_{L^2 \setminus \{0\}} F = \inf_{\sphereL} F|_{\sphereL}$ and $l^\star$ minimizes $F$ if
and only if $\hat l^\star$ minimizes $F|_{\sphereL}$. The constraint may thus be dropped, and the objective minimized over the whole of
$L^2([0,1],\mathbb{R})$ by a standard Euclidean method.
In fact, the puncture is immaterial for the optimization: $F$ is constant along rays, so its
gradient is orthogonal to the current iterate and no descent direction points towards
the origin, and in practice the norm of the iterates stays within a few percent of its
starting value.

The gradient needed by such a method is moreover the one already at hand, up to a renormalization. By Euler's
identity, a differentiable function homogeneous of degree zero satisfies
$\langle \nabla F(l), l \rangle_{L^2} = 0$ for every $l \neq 0$, so $\nabla F(l)$ is tangent
to $\sphereL$ at $\hat l$. Differentiating $F(c\,l) = F(l)$ in $c$ gives
\begin{align*}
  \nabla F(l) \;=\; \frac{1}{\|l\|_{L^2}}\, \nabla F(\hat l)
              \;=\; \frac{1}{\|l\|_{L^2}}\, P_{\hat l}\big(\nabla F(\hat l)\big)
              \;=\; \frac{1}{\|l\|_{L^2}}\, \mathrm{grad}\, F|_{\sphereL}(\hat l),
  \qquad l \neq 0,
\end{align*}
the middle equality because $\nabla F(\hat l)$ is already tangent, the last because $F$ is a
smooth extension of $F|_{\sphereL}$ \citep[][eq.~(3.37)]{absil2008optimization}. A single
implementation thus serves all three schemes, the two Riemannian ones evaluating it only at
$\|l\|_{L^2} = 1$. The new scheme is obtained by handing $\nabla F$ to the limited-memory
BFGS algorithm of \citet{byrd1995limited}, which avoids the retractions and vector transports
that a genuine Riemannian BFGS requires.

The two facts used above, that a degree-zero homogeneous objective has
the same minimizers on the sphere and on the whole space and that its gradient is orthogonal
to its argument, are familiar from the analysis of scale-invariant parameters in
normalized neural networks \citep{arora2019auto}. However, we are not aware of the device having been used for the optimization over
$\Gamma$. In the registration of two scalar functions, \citet{wu2026sobolev} also removes the
constraints on $\gamma$ by a change of variables, the centered log-ratio transform
of $\log \dot\gamma$, mapping $\Gamma$ onto a linear space, where the optimization is over a finite
basis under a Sobolev penalty. 

\subsection{Numerical comparison of the reparametrization methods}
\label{app:reparam_simulation}

We compare the three variants above with the two dynamic programming routines of \texttt{fdasrvf}: the classical one (\texttt{DPQ}), which searches over monotone piecewise-linear paths on a uniform $n \times n$ grid, and a variant (\texttt{DPQ2}) which decouples the grid on which the path is sought from the grid on which the curves are sampled, allows both to be non-uniform. Let $\beta$ be a fixed curve and $\gamma \in \Gamma$ a reparametrization. If $q_1$ and $q_2$ denote the SRVFs of $\beta \circ \gamma$ and of $\beta$ respectively, then $q_1 = (q_2 \circ \gamma)\sqrt{\dot\gamma}$. With the rotation fixed to the identity, the infimum of $H(\,\cdot\,, O)$ is therefore attained at $\gamma$ itself, and the retrieved warping can be compared with a target that is exact rather than itself the output of a numerical scheme.

As the curve $\beta$ we take a tongue contour from the rt-MRI recording used in the speech inversion application of the main paper, namely a frame whose contour is typical among those that are complete, smoothed as described in Appendix~\ref{app:contour_ordering} and evaluated on $n=100$ points, the resolution used in both applications. We draw $m = 200$ random monotone reparametrizations and record, for every warping and every method, the time per alignment and the error $\|\hat\gamma - \gamma\|_{L^2}$. We report that error rather than the geodesic distance on $\sphereL$ because it is defined for every method without further processing: measuring on $\sphereL$ would require differentiating $\hat\gamma$, which is natural for the methods that optimize over $\psi = \sqrt{\dot\gamma}$ but artificial for the dynamic programming routines, whose $\hat\gamma$ is piecewise linear on a grid by construction.

\begin{figure}[p]
  \centering
  \includegraphics[width=.48\textwidth]{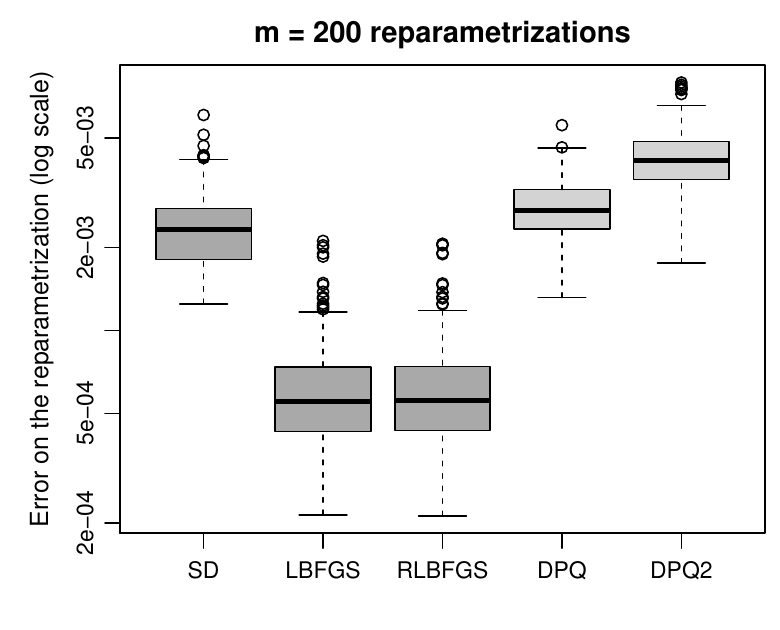}
  \hfill
  \includegraphics[width=.48\textwidth]{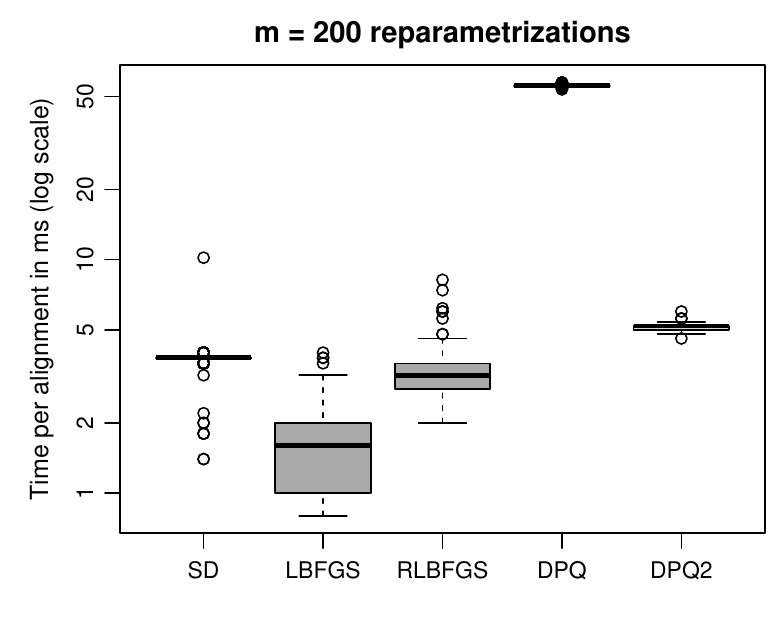}
  \caption{Retrieval of a known reparametrization of a tongue contour discretized on
  $n=100$ points, over $m=200$ random monotone reparametrizations. Left: the error
  $\|\hat\gamma-\gamma\|_{L^2}$, right: time per alignment. Both axes are logarithmic. In dark grey our three
  schemes for the optimization over $\Gamma$: the steepest descent of
  Algorithm~\ref{code:steepdesc} (SD), the quasi-Newton method applied to the
  scale-invariant extension of the cost (LBFGS), and a Riemannian quasi-Newton method
  on $\sphereL$ itself (RLBFGS). In light grey the two dynamic programming routines of
  \texttt{fdasrvf} \citep{tucker2025fdasrvf}, the classical one (DPQ) and its variant
  on a decoupled, possibly non-uniform grid (DPQ2), which are the only implementations available in
  \texttt{R} for planar curves.}
  \label{fig:methods_alignment}
\end{figure}

The results are reported in Figure~\ref{fig:methods_alignment}. The two quasi-Newton
schemes are the most accurate by a factor of at least four, with boxes that do not overlap
with any of the others, and the \texttt{LBFGS} variant is at the same time the fastest of
all the methods. The computational cost
separates the methods in the same direction.

The two quasi-Newton schemes agree to within one percent, and even though the Riemannian scheme needs the fewest gradient evaluations of all the methods considered, 
it is twice as slow, due to the repeated computation of inner products in $L^2$.

\begin{algorithm}[p] \caption{Riemannian steepest descent algorithm} \label{code:steepdesc} 
\begin{algorithmic}[1] 
\State Given the cost function $C:\sphereL\rightarrow \mathbb{R}$, a starting point $l_0$ and a tolerance $\epsilon>0$: 
\For{$k = 0, 1, \dots$} \State Compute $\nabla C(l_k)$ 
\If{$\norm{\nabla C(l_k)}<\epsilon$} 
\State \textbf{return} $l_k$ 
\EndIf 
\State Compute a line-search step size $\alpha_k$ using Armijo rule (see Algorithm \ref{code:armijo}) 
\State Set $l_{k+1}= \Exp_{l_k}(- \alpha_k \nabla C(l_k))$ 
\EndFor 
\end{algorithmic} \end{algorithm}

\begin{algorithm}[p]
  \caption{Armijo backtracking line search on $\sphereL$}
  \label{code:armijo}
  \begin{algorithmic}[1]
  \State Given the cost $C:\sphereL\rightarrow \mathbb{R}$, the current point
    $l\in \sphereL$ with Riemannian gradient $g = \operatorname{grad} C(l)$,
    constants $\tau \in [10^{-5}, 10^{-1}]$ and $\eta \in [1/10, 1/2]$.
  \State Set $m = 0$, $\alpha = 1$, and $l^{\text{new}} = \Exp_{l}(-\alpha\, g)$.
  \While{$C(l)-C(l^{\text{new}}) < \tau\, \alpha\, \langle g, g\rangle_{L^2}$}
    \State $m \gets m+1$; \quad $\alpha \gets \eta^{m}$
    \State $l^{\text{new}} \gets \Exp_{l}(-\alpha\, g)$
  \EndWhile
  \State \textbf{return} $\alpha$
\end{algorithmic}
\end{algorithm}

\section{Simulation: data generation}
\label{app:simulation}

\begin{figure}[H]
\centering
  \includegraphics[width=.7\textwidth]{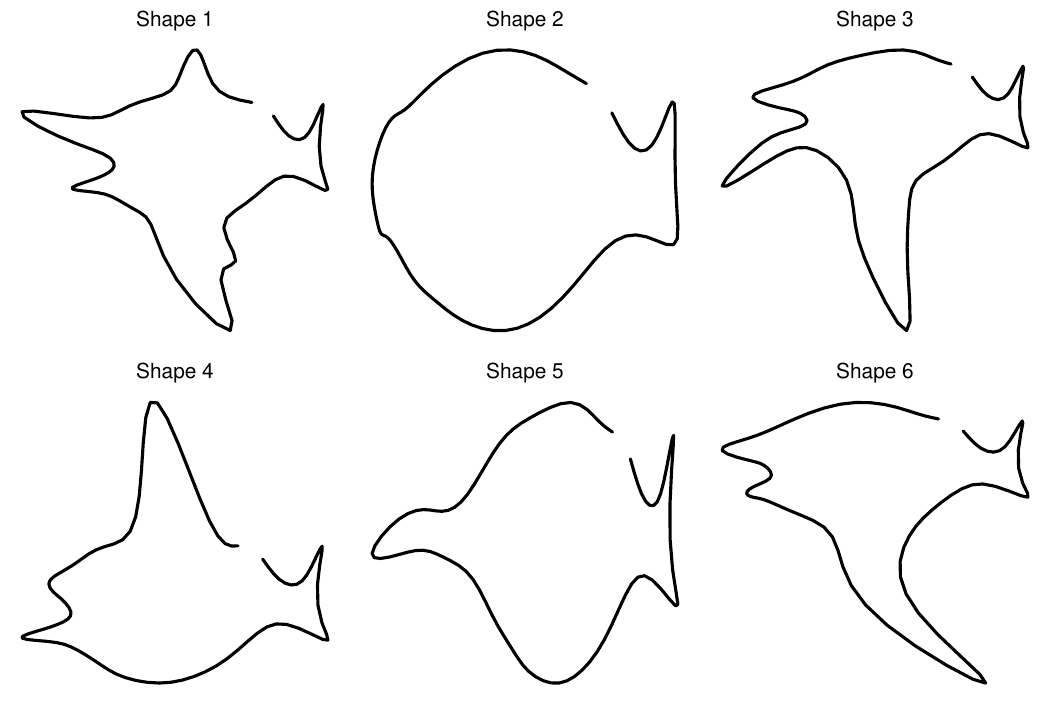}
  \caption{The six hand-drawn fish shapes from which the two generating trajectories are built. Shapes $1$ to $3$ (top row) and $4$ to $6$ (bottom row) are placed at $x = -1$, $0$ and $1$ respectively, from left to right.}
  \label{fig:6_fishes}
\end{figure}

\begin{figure}[H]
\centering
  \includegraphics[width=.6\textwidth]{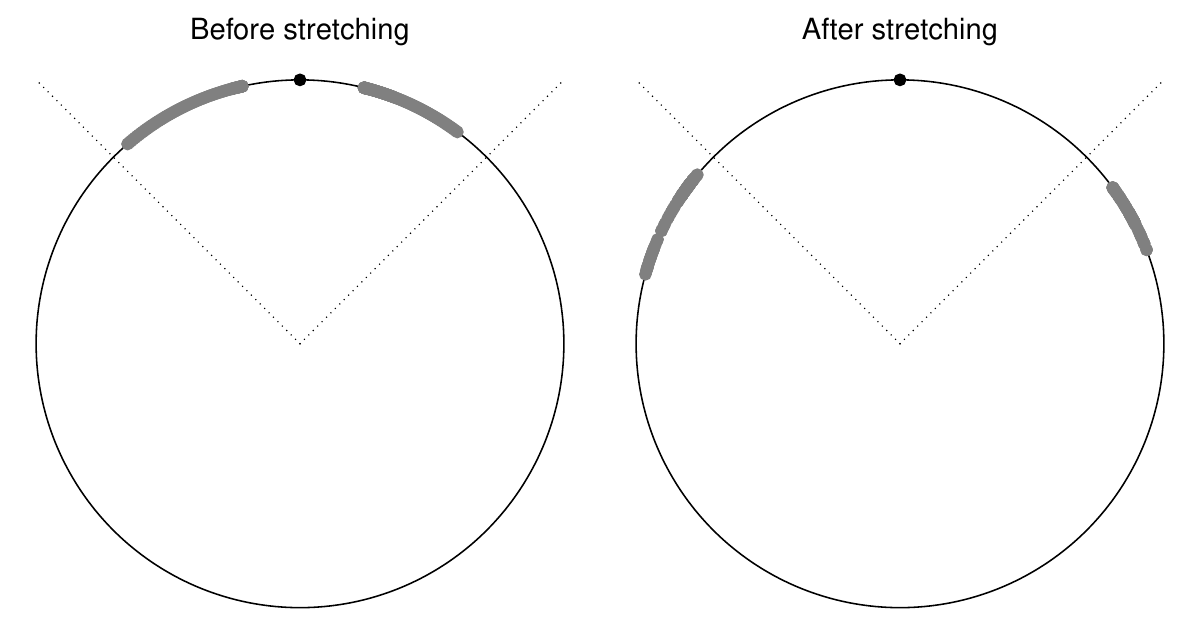}
  \caption{Shapes $1$, $3$, $4$ and $6$ (Figure~\ref{fig:6_fishes}) are stretched along the geodesic segments joining them to shapes $2$ and $5$, to produce data more spread out on the sphere.}
  \label{fig:data_spread}
\end{figure}

\begin{figure}[H]
\centering
  \includegraphics[width=.3\textwidth]{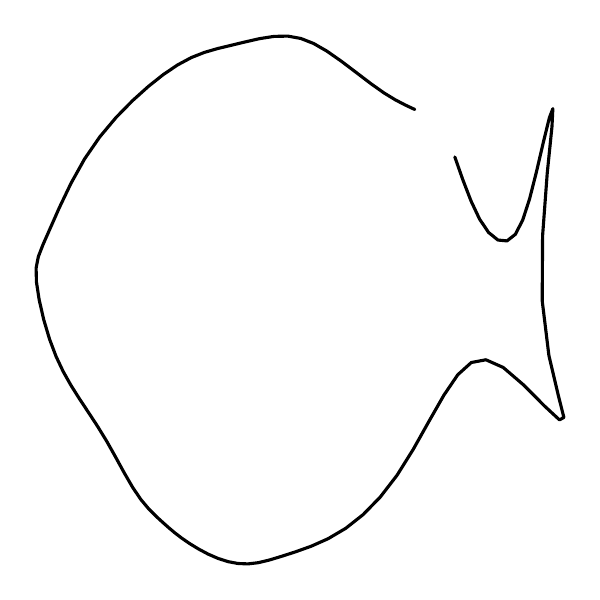}
  \caption{The template shape $\origin$ at which the two generating regressions are based; it is distinct from the six shapes above.}
  \label{fig:origin_fish}
\end{figure}

The ground truth of the simulation study is built from six fish outlines drawn
by hand (Figure~\ref{fig:6_fishes}), smoothed and mapped to unit-norm SRVFs
on $100$ points, and from a seventh outline, the template $\origin$ of
Figure~\ref{fig:origin_fish}, which is the base point of every regression
of the study, both of the two generating regressions below and of the models
fitted to the simulated samples. Shapes $1$ to $3$ give the trajectory
$\eta_0$ of Section~5.2 of the main paper and shapes $4$ to $6$ the trajectory
$\eta_1$: each triplet is placed at the covariate values $x = -1, 0, 1$ and
fitted by an elastic spherical regression with base point $\origin$, a GRBF
kernel with bandwidth $\sigma^2 = 1$ (the median squared distance between the
three covariates) and $\lambda = 0.01$, and the fitted trajectory is evaluated
on a grid of $200$ equispaced covariate values, on which the true conditional
means are defined, from which the samples draw their covariates and on which
the error is scored. Before fitting, the shapes at the ends of
each triplet are stretched: shape $1$ is moved along the geodesic from shape
$2$ through shape $1$ to $2.5$ times its distance from shape $2$, and shapes
$3$, $4$ and $6$ likewise with respect to shapes $2$ and $5$. Without the
stretching the six shapes lie close to one another on the sphere and the
simulated responses would stay in a neighbourhood of the base point. Figure~\ref{fig:data_spread} shows
the spread of the trajectory on the sphere before and after it. Finally, at every $x$ the shape
$\eta_1(x)$ is aligned to $\eta_0(x)$ over rotation and reparametrization,
and the true conditional mean $\mu(x)$ is the midpoint of the geodesic joining
them, as in Section~5.2.

\section{Data preparation for the temporal smoothing of vocal tract shapes}
\label{app:contour_ordering}

This section details how the vocal tract shapes of the temporal smoothing application are obtained from the segmentation masks of the rt-MRI frames: the labeled pixels of each articulator are first ordered along its contour, and the ordered contours are then joined into a single open curve.

The labeled pixels are taken as the vertices of a complete graph with Euclidean edge weights and we compute its minimum spanning tree \citep{kruskal1956shortest,prim1957shortest} and extract the diameter of the tree, i.e.\ its longest weighted simple path. Because a spanning tree connects all points by construction, gaps of any width in the pixel contour are bridged by a single edge, the only constant retained is a threshold discarding isolated pixels (3 px) before the tree is built. Isolated misclassified pixels and small side branches become short spurs of the tree, which the diameter path does not visit. Then, the diameter is a simple path running from one end of the contour to the other, so it directly provides the ordering sought. The use of the Euclidean minimum spanning tree to order a sample of points along a plane curve goes back to \citet{figueiredo1995computational}, who show that for a sufficiently dense sample of an arc the tree is a path.

The vocal tract contour used in the temporal smoothing application is assembled from seven articulators labeled by the segmentation: three along the upper boundary of the tract (the upper lip, the hard palate and the velum, from front to back) and four along its lower boundary (the epiglottis, the tongue, the lower incisors and the lower lip, from back to front). Each articulator is ordered separately as described above. The contours are then oriented consecutively (the first so that it runs from the lips backwards, each following one so that it starts at the end closer to where the previous one ends) and concatenated, the upper boundary from front to back followed by the lower boundary from back to front, so that the two ends of the resulting open curve are the two lips. After smoothing, every curve is given the same direction of traversal, starting at the lower lip: a reversed curve would be a different SRVF, at a large elastic distance from its neighbours, since reparametrizations preserve the direction.

The two boundaries do not meet in the image: between the back of the velum and the top of the epiglottis lies the oropharynx. We bridge this gap with a straight segment of $15$ equally spaced points, inserted before the smoothing spline so that the smoothed curve passes continuously from one boundary to the other. This connector has no anatomical meaning. It is a formal device that lets the whole vocal tract be represented as a single open curve. Since the SRVF representation used in the main paper discards translation and scale, representing the two boundaries as separate curves would lose their relative position and size, and with them the separation between the two parts of the tract, whereas as a single curve they share one coordinate frame, and the shape retains the correct proportions between them. Being part of the representation rather than of the anatomy, the connector is not drawn in the figures, which show the two boundaries only. The tongue contours of the speech inversion application consist of a single articulator and need no connector.

\section{Speech signal and MFCCs}

Sound is a longitudinal air–pressure wave recorded by a microphone at a fixed rate, 
which we represent as a discrete time series $s(t)$, 
for $t = 1,\dots,T$. To analyze how its frequency content evolves over time, we first 
compute a spectrogram, that is, a time–frequency representation obtained by applying 
a short‑time Fourier transform (STFT) with a symmetric window function 
$w \colon \mathbb{R} \to \mathbb{R}$ to overlapping 
frames of $s(t)$. For a given time index $t$ and angular frequency $\omega$, the STFT is
\begin{align*}
\mathrm{STFT}(s)(t,\omega)
  &= \sum_{u=-\infty}^{\infty} s(u)\, w(u-t)\, \exp(-i\omega u) = \sum_{u=1}^{T} s(u)\, w(u-t)\, \exp(-i\omega u),
\end{align*}
where $u$ indexes the samples within the recording and it is understood that $s(u) = 0$ 
for $u \le 0$ or $u > T$. The corresponding spectrogram is the squared magnitude of the 
STFT,
\begin{equation*}
  \mathrm{Spec}(s)(t,\omega)
    = \bigl|\mathrm{STFT}(s)(t,\omega)\bigr|^2 ,
\end{equation*}
which describes how the signal energy is distributed across time $t$ and frequency 
$\omega$.

Mel‑frequency cepstral coefficients (MFCCs) provide a lower‑dimensional summary of this 
spectrogram that reflects properties of human hearing. To construct them, first the mel-spectrogram is computed
\begin{equation*}
\mathrm{MelSpec}(s)(t,f)
  = \sum_{k=0}^{N-1} \mathrm{Spec}(s)\bigl(t, 2\pi k / N\bigr)\, b_{f,k},
\end{equation*}
where $N\geq T$ is a power of $2$ and 
$(b_{f,k})_{k=0}^{N-1}$, for $f = 1, \dots, F$, is the Mel-scale filter bank with $F$ filters, which is believed to approximate the functioning of the human auditory system. The MFCCs are then obtained by:
\begin{equation*}
\mathrm{MFCC}(s)(t,m)
  = \frac{1}{F} \sum_{f=0}^{F}
      \log\!\bigl(\mathrm{MelSpec}(s)(t,f)\bigr)
      \exp\!\left( i\,\frac{2\pi (m-1) f}{F+1} \right),
\end{equation*}
for $ m = 1,\dots,M $. For a more detailed description of spectrograms and MFCCs see, e.g., \citet{koshy2022exploring} and \citet{tavakoli2019spatial}.

In this application, before extracting MFCCs from the raw speech recordings, we perform a simple denoising step in Audacity \citep{audacity}. We apply a high‑pass and a low-pass filter with cutoff 1000 Hz and a 6 dB‑per‑octave roll‑off to suppress background noise, and then use the resulting cleaned waveform as input to the MFCC computation.

Our construction of the acoustic covariates in this implementation (including the choice of all the hyper-parameters) follows \citet{liu2015deep} and \citet{azzouz2025complete}.
To compute MFCCs we apply the function \texttt{melfcc} from the \texttt{tuneR} package, which returns the first $M = 13$  cepstral coefficients $\mathbf{c}_t \in \mathbb{R}^{13}$, at each time frame $t$. The MFCCs are
calculated with a window length of 25 milliseconds and a frame shift of
10 milliseconds.

Following common practice in phonetics, we also include first‑ and
second‑order temporal derivatives of the MFCC trajectories, often called
“deltas” $\boldsymbol{\Delta}\mathbf{c}_t, \in \mathbb{R}^{13}$ and “double–deltas” $\boldsymbol{\Delta}^2\mathbf{c}_t \in \mathbb{R}^{13}$, computed with the \texttt{deltas} function
in \texttt{tuneR}. These approximate the local velocity and acceleration of the MFCCs over time and are widely used because many phonetic contrasts and coarticulatory effects are encoded in how the spectrum changes over time rather than in its instantaneous profile alone.
To ensure that MFCCs and their temporal derivatives contribute equally
in subsequent analyses, each block
$\mathbf{c}_t, \boldsymbol{\Delta}\mathbf{c}_t, \boldsymbol{\Delta}^2\mathbf{c}_t$ is
separately standardized and then normalized so that its total variance
is one. We then form a 39‑dimensional feature vector
\[
\mathbf{f}_t =
\bigl(\mathbf{c}_t^\top,\ \boldsymbol{\Delta}\mathbf{c}_t^\top,\ \boldsymbol{\Delta}^2\mathbf{c}_t^\top\bigr)^\top
\in \mathbb{R}^{39},
\]
and apply a temporal context window of 11 frames, consisting of the
current frame, the five preceding frames, and the five subsequent
frames, to obtain
\[
\mathbf{x}_t =
\bigl(\mathbf{f}_{t-5}^\top,\dots,\mathbf{f}_t^\top,\dots,\mathbf{f}_{t+5}^\top\bigr)^\top
\in \mathbb{R}^{429}.
\]

Finally, the MFCC-based feature vectors are temporally aligned to the
video frame rate of 83.28 frames per second via linear interpolation,
yielding a matrix
\(
X = (\mathbf{x}_1^\top,\dots,\mathbf{x}_T^\top)^\top \in \mathbb{R}^{T \times 429}
\),
with one acoustic covariate vector $\mathbf{x}_t$ per video frame $t$.

Silent frames were detected based on RMS energy computed over each video frame \citep{asadullah2016silence}, and frames corresponding to background silence or missing tongue-shape data were removed. Principal Component Analysis (PCA) is subsequently applied to the resulting feature matrix in order to reduce its dimensionality. The number of retained components was set to $20$, which explain $94\%$ of the variance of the features on the training frames.

\section{Speech inversion: shape error in millimetres}
\label{app:si_mm}

The error in millimetres quoted in Section~7 of the main paper is computed
from the stored predictions as follows. For each of the $240$ held-out
frames, the predicted SRVF $\hat q$ is mapped back to a curve by integrating
$\hat q\,|\hat q|$ on the grid. This curve starts at the origin, has unit
length and carries no orientation of its own, since position, size and
rotation are precisely what the SRVF representation removes. The observed
tongue of the corresponding frame is the smoothed contour in the $84\times 84$ pixel
grid from which the held-out SRVF was computed. Both curves are resampled at
$n = 50$ points equally spaced along their length, as many as the tracker of
\citet{azzouz2026embeddings} returns, and matched by index, so that no reparametrization
enters this computation, unlike in the elastic distance of the main text.
The predicted points $c_j$ are then carried onto the observed ones $b_j$ by
the scaling $s>0$, rotation $R \in SO(2)$ and translation $t \in \Real^2$
that minimise $\sum_{j=1}^{n} \|s R c_j + t - b_j\|^2$, an ordinary full
Procrustes superimposition, whose scaling replaces the unit length of the
reconstructed curve by the length of the tongue in that frame. Writing
$(x_j, y_j)$ for the observed points and $(\hat x_j, \hat y_j)$ for the
placed predictions, both in pixels, the error of the frame is
\[
  \mathrm{RMSE} \;=\; \sqrt{\frac{1}{2n}\sum_{j=1}^{n}
  \Big[(x_j-\hat x_j)^2 + (y_j-\hat y_j)^2\Big]}\,,
\]
the root mean square of the $2n = 100$ coordinate differences, multiplied by
the pixel size of $2.4$ mm of the recording
\citep{somandepalliSemanticEdgeDetection2017}. This is the formula of
\citet{azzouz2026embeddings}, applied to the vector of
the $100$ coordinates of each contour. Averaged over the held-out frames it gives $1.9$ mm 
for \ourMethod\ and $2.6$ mm
for the Fr\'echet mean of the training shapes. 

This number is not directly comparable with those of 
\citet{azzouz2026embeddings}, for several
reasons. First, our procedure hands the prediction the position, size and
rotation, of the observed tongue, whereas their network predicts the
coordinates in the image and is charged for the errors in these. Their task
is harder in this respect, and the difference is not small: given the pose,
even the Fr\'echet mean of the training shapes, which uses no acoustic
information at all, falls within their range. The difference reflects the
two formulations rather than the evaluation. In the SRVF framework the shape
is the object of the regression and the pose is removed by construction, so
that the accuracy of the shape can be assessed on its own. The position, size and rotation alignment used
here come from the observed tongue, which is available for evaluation but
not when the model is used to predict. 
This
decomposition suits the present purpose, where the shape of the articulator
is the quantity of interest, but when the whole configuration of the vocal
tract is sought, position, size and rotation carry articulatory information of their
own, and predicting them is part of the task. If position, size and
rotation were to be made part of the task in our model, as in \citet{azzouz2026embeddings}, they would have to be modelled separately, as a regression of a few
numbers, rather than learned together with the shape from the image
coordinates. Second, the
corpora differ by orders of magnitude: about ten seconds of training speech,
three vowels around one consonant, tested on a held-out section containing an
unseen consonant, against $2100$ sentences and $3.5$ hours tested on a random
$10\%$ of the acquisitions. Third, the speakers and the images differ. Their
speaker is female and ours male, so that for the same relative accuracy the
error in millimetres is expected to be larger on the longer male vocal tract
(our tongue contour is about $141$ mm long, theirs is not reported). 
The ground truth of either study is a contour extracted from the image, and
is therefore only known to about a pixel, which is $1.62$ mm in their images
and $2.4$ mm in ours. The contours have different extents, from the
tongue root to the sublingual cavity in their tracker and along the pixels
labelled as tongue by the segmentation in ours. Finally, neither
their code nor their corpus is released, so that none of these differences
can be removed by running either method on the other data.

\bibliographystyle{agsm} 
\bibliography{elastic-bib-shahin,elastic-bib-bea,extra-elastic}
 
\label{lastpage}

\end{document}